\documentclass[fleqn,usenatbib]{mnras}
\usepackage{amsmath}	% Advanced maths commands
\usepackage{amssymb}	% Extra maths symbols
\usepackage{newtxtext,newtxmath}
\usepackage[normalem]{ulem}

\usepackage[T1]{fontenc}
\DeclareRobustCommand{\VAN}[3]{#2}
\let\VANthebibliography\thebibliography
\def\thebibliography{\DeclareRobustCommand{\VAN}[3]{##3}\VANthebibliography}

\usepackage{graphicx}	% Including figure files
\usepackage{hyperref}
\usepackage{color}
\usepackage{multirow}
\usepackage{verbatim}
\usepackage{tabularx}
\usepackage{xspace}	
\usepackage{xcolor}
\usepackage{sistyle}
\usepackage{orcidlink}
\SIthousandsep{,}
\graphicspath{{./}{figures/}}

\newcommand{\unit}[1]{\ensuremath{\mathrm{\,#1}}\xspace}
\newcommand{\feh}{\mathrm{[Fe/H]}}

\newcommand{\logg}{\ensuremath{\log\,g}\xspace}

\newcommand{\etot}{E_{\rm tot}}
\newcommand{\lz}{L_{\rm z}}
\newcommand{\vr}{V_{R}}
\newcommand{\vphi}{V_{\phi}}
\newcommand{\vz}{V_{z}}
\newcommand{\jr}{J_r}
\newcommand{\jphi}{J_{\phi}}
\newcommand{\jz}{J_z}
\newcommand{\zmax}{z_{\mathrm{max}}}
\newcommand{\rperi}{r_{\mathrm{peri}}}
\newcommand{\rapo}{r_{\mathrm{apo}}}

\newcommand{\kms}{\unit{km\,s^{-1}}}
\newcommand{\kmkmss}{\unit{km^{2}\,s^{-2}}}
\newcommand{\kpckms}{\unit{kpc\,km\,s^{-1}}}
\defcitealias{bailer-jones:21}{BJ21}

\DeclareUnicodeCharacter{2212}{-}%%%%%%%%%%%%%%%%%%%%%%%%%%%%%%%%%%%%%%%%%%%%%%%%%%

\title[Nearby Halo Substructures in DESI MWS Y1]{Nearby stellar substructures in the Galactic halo from DESI Milky Way Survey Year 1 Data Release}
\author[Kim et al.]{Bokyoung Kim\orcidlink{0000-0002-8999-1108}$^{1}$\thanks{E-mail: bokyoung.kim@ed.ac.uk},
Sergey E. Koposov\orcidlink{0000-0003-2644-135X}$^{1,2}$
Ting S. Li\orcidlink{0000-0002-9110-6163}$^{3}$, 
Sophia Lilleengen\orcidlink{0000-0001-9046-691X}$^{4}$,
Andrew P. Cooper\orcidlink{0000-0001-8274-158X}$^{5,6}$,
\newauthor
Andreia Carrillo\orcidlink{0000-0002-5786-0787}$^{7,4}$,
Monica Valluri\orcidlink{0000-0002-6257-2341}$^{8}$,
Alexander H.~Riley\orcidlink{0000-0001-5805-5766}$^{4}$,
Jiwon Jesse Han\orcidlink{0000-0002-6800-5778}$^{9}$,
\newauthor
Jessica Nicole Aguilar$^{10}$,
Steven Ahlen\orcidlink{0000-0001-6098-7247}$^{11}$,
Leandro~{Beraldo e Silva}\orcidlink{0000-0002-0740-1507}$^{8,12}$,
Davide Bianchi\orcidlink{0000-0001-9712-0006}$^{13}$,
\newauthor
David Brooks$^{14}$,
Amanda Bystr\"om$^{1}$, 
Todd Claybaugh$^{10}$,
Shaun Cole\orcidlink{0000-0002-5954-7903}$^{4}$,
Kyle Dawson$^{15}$,
\newauthor
Axel de la Macorra\orcidlink{0000-0002-1769-1640}$^{16}$,
Jaime Forero-Romero\orcidlink{0000-0002-2890-3725}$^{17,18}$,
Oleg Y. Gnedin\orcidlink{0000-0001-9852-9954}$^{19}$,
Satya Gontcho A Gontcho\orcidlink{0000-0003-3142-233X}$^{10}$,
\newauthor
Gaston Gutierrez$^{20}$,
Julien Guy\orcidlink{0000-0001-9822-6793}$^{10}$,
Klaus Honscheid$^{21,22,23}$,
Robert Kehoe$^{24}$,
Namitha Kizhuprakkat$^{5,6}$,
\newauthor
Martin Landriau\orcidlink{0000-0003-1838-8528}$^{10}$,
Laurent Le Guillou\orcidlink{0000-0001-7178-8868}$^{25}$,
Michael Levi\orcidlink{0000-0003-1887-1018}$^{10}$,
Gustavo E. Medina\orcidlink{0000-0003-0105-9576}$^{3}$,
\newauthor
Aaron Meisner\orcidlink{0000-0002-1125-7384}$^{26}$,
Ramon Miquel$^{27,28}$,
John Moustakas\orcidlink{0000-0002-2733-4559}$^{29}$,
Claire Poppett$^{10,30,31}$,
Francisco Prada\orcidlink{0000-0001-7145-8674}$^{32}$,
\newauthor
Graziano Rossi$^{33}$,
Eusebio S\'{a}nchez\orcidlink{0000-0002-9646-8198}$^{34}$,
Michael Schubnell$^{35,19}$,
Ray Sharples\orcidlink{0000-0003-3449-8583}$^{36,4}$,
David Sprayberry$^{26}$,
\newauthor
José Arturo Trelles Hernández$^{16}$,
Benjamin Alan Weaver$^{26}$,
and Hu Zou\orcidlink{0000-0002-6684-3997}$^{37}$\\
Authors affiliations are provided in the Appendix \ref{sec:affiliations}
}
\date{Accepted 2025 April 28. Received 2025 April 8; in original form 2024 October 2}

\pubyear{2024}

\begin{document}
\label{firstpage}
\pagerange{\pageref{firstpage}--\pageref{lastpage}}
\maketitle

\begin{abstract}
We report five nearby ($d_{\mathrm{helio}} < 5$ kpc) stellar substructures in the Galactic halo from a subset of \num{138661} stars in the Dark Energy Spectroscopic Instrument (DESI) Milky Way Survey Year 1 Data Release. With an unsupervised clustering algorithm, HDBSCAN*, these substructures are independently identified in Integrals of Motion ($\etot$, $\lz$, $\log{\jr}$, $\log{\jz}$) space and Galactocentric cylindrical velocity space ($\vr$, $\vphi$, $\vz$). We associate all identified clusters with known nearby substructures (Helmi streams, M18-Cand10/MMH-1, Sequoia, Antaeus, and ED-2) previously reported in various studies. With metallicities precisely measured by DESI, we confirm that the Helmi streams, M18-Cand10, and ED-2 are chemically distinct from local halo stars. We have characterised the chemodynamic properties of each dynamic group, including their metallicity dispersions, to associate them with their progenitor types (globular cluster or dwarf galaxy). Our approach for searching substructures with HDBSCAN* reliably detects real substructures in the Galactic halo, suggesting that applying the same method can lead to the discovery of new substructures in future DESI data. With more stars from future DESI data releases and improved astrometry from the upcoming {\it Gaia} Data Release 4, we will have a more detailed blueprint of the Galactic halo, offering a significant improvement in our understanding of the formation and evolutionary history of the Milky Way Galaxy.
\end{abstract}


%%%%%%%%%%%%%%%%%%%%%%%%%%%%%%%%%%%%%%%%%%%%%%%%%%

%%%%%%%%%%%%%%%%% BODY OF PAPER %%%%%%%%%%%%%%%%%%

\section{Introduction}
The Milky Way halo is a laboratory for Galactic archaeologists who conduct observational and computational experiments to understand the formation and evolution history of the Milky Way. The halo population consists of old stars with relatively chemically pristine atmospheres that conserve information about the early Universe. Halo stars are, thus, great tracers for understanding the ancient formation and accretion history, as well as the chemical enrichment history of the Milky Way \citep{eggen:62, sz:78, fuhrmann:98, freeman:02, carollo:07}.

While some illustrations of the Milky Way picture the Galactic halo as a smooth, homogeneous sphere, these figures miss the main features of the Milky Way halo regarding a number of stellar substructures and overdensities. Stellar substructures are tidal debris originating from star clusters or dwarf galaxies \citep{newberg:16, helmi:20, bonaca:24}, with members of each substructure sharing similar chemodynamic characteristics. Substructures originating from globular clusters appear as thin, dynamically cold stellar streams due to small intrinsic velocity dispersions of their progenitors, whereas tidal streams produced by accreted dwarf galaxies show more dispersed spatial distributions. Furthermore, tidal debris from both globular clusters and dwarf galaxies have unique elemental abundance signatures, as the chemical signatures of stars preserve the environmental information of the molecular clouds where stars formed \citep{freeman:02, desilva:15}. Stellar debris with different origins, thus, show unique distributions in chemodynamic parameter spaces, which provides information allowing us to trace the origin of these structures \citep[see e.g.,][]{helmi:20, deason:24, bonaca:24}. 

The existence of stellar substructures in the Milky Way has been known for over three decades. The very first discovery of the Sagittarius dwarf galaxy \citep{ibata:94} supported the idea that the Milky Way is currently experiencing an ongoing merger event. \citet{helmi:99} reported a heavily phase-mixed substructure in the Solar neighbourhood with precisely measured distances and proper motions of stars with the {\it Hipparcos} \citep{perryman:97}. The Sloan Digital Sky Survey \citep[SDSS;][]{york:00} and the Two Micron All Sky Survey \citep[2MASS;][]{skrutskie:06} are pioneers that expanded sample size and reported dozens of stellar streams in the outer halo \citep[e.g.,][]{odenkirchen:01, rockosi:02, majewski:03, odenkirchen:03, belokurov:06a, belokurov:06b, grillmair:06a, grillmair:06b, bonaca:12}. Later, the {\it Gaia} mission \citep{gaia:16, gaia:18, gaia:21} revolutionised the whole search for stellar substructures by providing a high-quality astrometric dataset for a few billion stars within a few tens of kiloparsecs from the vicinity of the Sun. One of the main discoveries from {\it Gaia} is the discovery of {\it Gaia-Sausage/Enceladus} \citep[GSE;][]{belokurov:18, helmi:18, haywood:18}. Since {\it Gaia} Data Release 2, more than $100$ stellar streams have been identified by various studies \citep{malhan:18a, myeong:18a, myeong:18b, yuan:18, koppelman:19b, myeong:19, naidu:20, horta:21, lovdal:22, malhan:22, ruiz-lara:22a, shank:22a, shank:22b, mikkola:23, mateu:23, bonaca:24}.

An obstacle to substructure studies is, however, the limited information on radial velocities for most stars in the Galactic halo. This is due to their faint brightness \citep[e.g., the magnitude limit of {\it Gaia} Radial Velocity Spectrometer, $G < 15$;][]{gaia:23b} and the cost of spectroscopic observation. These barriers have prevented detailed kinematic analysis of the Milky Way stellar halo, as well as limiting our understanding of the chemical properties of the majority of the halo population.

The Dark Energy Survey Instrument \citep[DESI;][]{desi:16a, desi:22, desi:23a, desi:23b} Milky Way Survey \citep[MWS;][]{allende_prieto:20, cooper:23, koposov:24} is designed to overcome this difficulty by providing highly accurate radial velocity and metallicity measurements for approximately seven million stars in the Milky Way halo. With mid-resolution spectra obtained from the Mayall 4-m telescope at Kitt Peak National Observatory, DESI MWS allows us to conduct detailed chemodynamic analysis on these objects. The application of the DESI MWS dataset spans a wide range of topics, from stellar population analysis to constraining the properties of the Milky Way dark matter halo.

In this work, we report our results of characterising stellar substructures in the local Galactic halo ($d < 5$ kpc) by performing an unsupervised machine learning clustering algorithm \citep[HDBSCAN*;][]{hdbscan} in the DESI MWS Year 1 Data Release (hereafter, the Y1 catalogue). By characterising substructures in the Solar vicinity, we aim to connect them to their origins with precisely measured DESI metallicities. A brief description of the DESI survey and the Y1 catalogue is provided in Section~\ref{sec:data}, as well as our sample selection process and the application of HDBSCAN*. In Section~\ref{sec:results}, we provide detailed information on substructures identified in this study and the validation process. In Section~\ref{sec:discussion}, we present the association of the identified substructures with well-known substructures or streams and discuss the potential progenitors of the substructures using DESI metallicities. We summarise our results in Section~\ref{sec:summary}.

\section{Data and Methods}\label{sec:data}
This work used the Y1 catalogue \citep[][I]{desi:24a} covering the first year of survey observation from DESI, which includes commissioning, survey validation, main survey, and special survey data. The Y1 key science papers present the two-point clustering measurements and validation \citep[][II]{desi:24b}. The DESI Y1 BAO measurements from galaxies and quasars \citep[][III]{desi:24c} and from the Ly$\alpha$ forest \citep[][IV]{desi:24d} play a key role in providing more precise measurements of the expansion rate of the Universe. Furthermore, the main Y1 science results include a full morphological study of galaxies and quasars \citep[][V]{desi:24e}, Cosmological results from the BAO measurements \citep[][VI]{desi:24f} and the full-shape analysis \citep[][VII]{desi:24g}, as well as constraints on primordial non-gaussianities (DESI Collaboration et al. 2024 VIII).

Here, we provide a summary of the DESI MWS Y1 catalogue and how we constructed a subset of stars in the local Galactic halo. More detailed information on DESI and MWS catalogues, including target selection, observations, data processing, and the MWS catalogue overview, is available in \citet{cooper:23, desi:23a, desi:23b, myers:23, guy:23, schlafly:23} and \citet{koposov:24}.

\subsection{DESI Observations and Pipelines}

DESI is an optical multi-object spectrograph with \num{5000} robotic positioners, the ends of which are connected to optical fibres \citep{desi:22}. The instrument is installed at the Mayall 4-m telescope, Kitt Peak National Observatory and covers over a $\sim 3^{\circ}$ field of view. It simultaneously obtains spectra of targeted objects \citep{desi:16b, silber:23, miller:23, poppett:24}. While the main goal of DESI is to investigate dark energy by precisely measuring the expansion rate of the Universe \citep{levi:13}, part of the five-year survey is also dedicated to observing over seven million unique stars in the Milky Way at high ($|b| > 20^{\circ}$) Galactic latitudes \citep{allende_prieto:20, cooper:23}. 

The Y1 catalogue includes data obtained from 14 December 2020 to 13 June 2022, also covering the Early Data Release Milky Way Survey Value Added Catalogue \citep[EDR MWS VAC;][]{koposov:24}. It contains over $6.16$ million optical spectra (\num{3600}\AA\ -- \num{9800}\AA) with mid-resolution ($R \sim$ \num{2000} -- \num{5000}) of stars having $r \le 19$ mag. All stars are targeted by using simple selection criteria for the thick disc and halo populations. The MWS Y1 catalogue has been crossmatched with the latest {\it Gaia}~data release 3 \citep{gaia:16, gaia:23a}, and all {\it Gaia} columns obtained from this process are included in FITS extensions. 

The stellar spectra observed by DESI were processed by {\tt RVSpecFit}\footnote{\url{https://github.com/segasai/rvspecfit}} \citep[RVS; ][]{koposov:19}, which fits interpolated stellar atmosphere models from the PHOENIX spectral library\footnote{\url{http://phoenix.astro.physik.uni-goettingen.de}} \citep{husser:13} to the blue, red, and infrared parts of the DESI spectra by performing the simultaneous, iterative least-squares optimisation. The outputs of the pipeline are maximum likelihood estimates of stellar atmospheric parameters, including radial velocity, stellar rotation velocity ($v\sin{i}$), and metallicity ($\feh$) \citep[for further details of the pipeline and its validation, see][]{cooper:23, koposov:24}. This work relies on the updated version of {\tt RVSpecFit} that uses a neural network-based interpolation of stellar templates. This version will also be used for the DESI DR1 release. The new interpolation technique in {\tt RVSpecFit} fixes the problem of gridding (i.e. concentration of stellar parameters near the grid points in temperature, metallicity and surface gravity) that was observed in the MWS EDR VAC \citep{koposov:24}. The other stellar pipeline, the Stellar Parameters (SP) pipeline using the {\tt FERRE} code\footnote{DESI-specific SP pipeline: \url{https://github.com/callendeprieto/piferre}} \citep{allende_prieto:08, cooper:23, koposov:24}, provides the main atmospheric parameters, such as effective temperature, surface gravity, microturbulence, metallicity, alpha-element abundance, and individual elemental abundances for C, Mg, Ca, and Fe \citep[more details are provided in][]{cooper:23, koposov:24}. In this work, we use the output from the RVS pipeline to conduct the chemodynamic analysis of nearby ($d_{\mathrm{helio}} < 5$ kpc) halo stars.

\subsection{DESI MWS Y1 catalogue}\label{sec:y1subset}
Following the recommendation for selecting clean stellar samples as outlined in \citet{koposov:24}, we applied initial selection criteria to select MWS targets by using bitmask fields described in \citet{myers:23} and removed non-stellar objects (i.e., galaxies or QSOs) and stars without {\it Gaia} DR3 information. Additionally, we implemented a stellar rotation velocity ($v\sin{i} < 50 \kms$) condition, primarily to filter out the majority of white dwarfs in the Y1 Catalogue, a radial velocity cut ($|V_{\mathrm{rad}}| < 600 \kms$) to remove stars with extreme radial velocities, and a condition for selecting stars with small uncertainties in their metallicity measurements from the RVS pipeline ($\sigma (\feh) < 0.2$). The selection criteria are as follows: 
\begin{itemize}
    \item {\tt DESI\_TARGET \& MWS\_ANY} $ > 0$\footnote{Selecting objects with the targeting bit ({\tt MWS\_ANY} = $2^{61}$), meaning that the objects were targeted for the Milky Way Survey \citep{cooper:23, myers:23, koposov:24}.}
    \item {\tt SV1/2/3\_DESI\_TARGET \& MWS\_ANY} $ > 0$
    \item {\tt RVS\_WARN = 0}\footnote{Selecting objects that {\tt RVSpecFit} obtained a good fit to the DESI spectrum.}
    \item {\tt RR\_SPECTYPE = STAR}
    \item $|V_{\mathrm{rad}}| < 600 \kms$
    \item $\sigma_{V_{\mathrm{rad}}} < 20 \kms$
    \item $-3.9 < \feh < 0.5$
    \item $\sigma_{\feh} < 0.2$
    \item {\tt source\_id} $> 0$\footnote{To remove objects that are not in {\it Gaia} DR3.}
    \item {\tt VSINI} $< 50 \kms$\footnote{Selecting objects with lower stellar rotational velocities.}.
\end{itemize}

Spectra of an object with multiple observations within a given survey and programme were coadded \citep[see also;][]{guy:23, koposov:24}. However, since spectra of objects that were observed multiple times across surveys and programmes are not coadded, these stars are reported multiple times in the Y1 dataset. We identified stars that were observed multiple times by searching for duplicated {\it Gaia} DR3 {\tt source\_id}. For these stars, we computed mean values of radial velocities and stellar parameters from multiple observations. 

\begin{figure}
    \centering
    \includegraphics{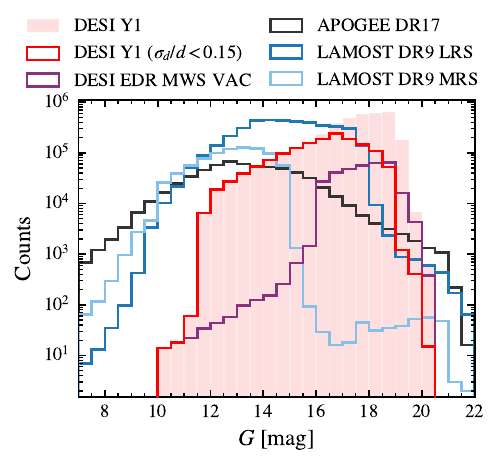}
    \caption{{\it Gaia} $G$ magnitude distribution of the Y1 catalogue (red-filled histogram: \num{3743580} stars), EDR MWS VAC subset (purple: \num{320918} stars) and external catalogues, including SDSS/APOGEE (black: \num{625393} stars) and LAMOST DR9 LRS (blue: \num{4080642} stars)/MRS (light blue: \num{830129} stars) catalogues. Stars with no {\it Gaia} G magnitude ($97$ stars in the Y1 subset and 18 stars in EDR MWS VAC) are not included, which is the reason that the number of stars in both the Y1 subset and EDR MWS VAC differs slightly from the reported values in the text. The red-unfilled histogram shows the same Y1 subset but only includes stars with good distance estimates ($\sigma_{d}/d < 0.15$: \num{1521249} stars).\label{fig:fig1}}
\end{figure}

The final number of stars in the Y1 catalogue (\num{3743677} stars, after applying the selection cuts and removing all duplicates) is more than 10 times larger than the EDR MWS VAC \citep[\num{320936} stars\footnote{Selected from the same selection criteria for constructing the Y1 catalogue.}; see also][]{koposov:24}. Figure~\ref{fig:fig1} shows the {\it Gaia} $G$ magnitude distribution of the Y1 catalogue after applying the selection criteria described above (red-filled histogram), compared to an equivalent selection from the EDR MWS VAC (purple histogram). In both selections, stars with no {\it Gaia} $G$ magnitude information ($97$ stars from Y1 and $18$ stars from EDR MWS VAC) are excluded. While both catalogues contain a significant number of faint objects ($G \ge 16$), about $39$ per cent of the Y1 catalogue consists of bright stars ($G < 16$) that were observed under poor observing conditions as part of the {\tt BACKUP} programme. At the fainter end ($G \ge 19.5$), on the other hand, EDR and Y1 show similar distributions. This is mainly because those stars were observed as part of survey validations with the majority of faint stars already included in the EDR dataset.

Figure~\ref{fig:fig1} also compares the {\it Gaia} $G$ magnitude distribution of the Y1 catalogue to other external spectroscopic surveys, including SDSS/APOGEE DR17 \citep[\num{625411} stars, black histogram;][]{majewski:17, abdurrouf:22}\footnote{\url{https://www.sdss4.org/dr17/irspec/}} and the LAMOST DR9 Low- and Mid-resolution stellar (LRS/MRS) parameter catalogues \citep[\num{4084621} stars from LRS shown as blue histogram, and \num{829991} stars from MRS shown as light blue histogram, respectively;][]{zhao:12, cui:12}\footnote{\url{https://www.lamost.org/dr9/}}. We removed duplicate sources from all external spectroscopy datasets, as well as non-stellar objects (i.e., {\tt STARFLAG} in APOGEE DR17), stars without a {\it Gaia} $G$ magnitude, and stars with mask bits indicating possible issues associated with data or problems with the stellar parameter pipeline fit (i.e., {\tt rv\_b\_flag} and {\tt rv\_r\_flag} in LAMOST DR9 MRS). We also applied a velocity error cut ($\sigma_{V_{\mathrm{rad, ext}}} < 20 \kms$) to the external spectroscopic surveys\footnote{For LAMOST DR9 MRS, we used radial velocity measurements errors in $r$ band, determined by using {\tt KURUCZ} \citep{kurucz:79, castelli:03} synthetic spectra ({\tt rv\_r0\_err}).}. 

The DESI Y1 catalogue provides high-precision radial velocities for fainter stars, which are unavailable in high-resolution spectroscopic surveys such as APOGEE. The DESI datasets (EDR and Y1) become even more valuable in $18 \le G < 20$, containing more observed stars than the other spectroscopic surveys. The number of stars in the Y1 catalogue drops at the very faint end (\num{642} stars with $G \ge 20$ shown in the red-shaded histogram in Figure~\ref{fig:fig1}) due to the limiting magnitude of DESI ($r < 20$). LAMOST DR9 has more stars with $G > 20$ (\num{1287} stars) compared to the Y1 dataset because the limiting magnitude of LAMOST ($r \sim 20.5$) is $0.5$ magnitude fainter than that of DESI.

To conduct kinematic analysis, we used 6-dimensional (6-D) phase-space data, i.e., positions and proper motions from {\it Gaia} DR3, distances from either {\it Gaia} parallaxes or \citet[][hereafter BJ21]{bailer-jones:21} and radial velocities from DESI. We used parallax distances for stars with precise parallax measurements (i.e., {\tt parallax\_over\_error} $> 10$) after applying a parallax zero-point correction \citep{lindegren:21a} and geometric distances from \citetalias{bailer-jones:21} for stars with {\tt parallax\_over\_error} $\le 10$. We computed uncertainties for the \citetalias{bailer-jones:21} distances by averaging the differences between the median distance and both the lower (16th percentile) and upper (84th percentile) distance estimates. Stars with distance uncertainties greater than $15$ per cent are excluded from the analysis. 

Our final dataset, combining all the criteria described above, contains \num{1521258} stars (hereafter, the Y1 subset; red-unfilled histogram in Figure~\ref{fig:fig1}), of which \num{6793} stars have heliocentric distances greater than $5$~kpc. The steep decline in the $G$ magnitude distribution of the Y1 subset around $G \sim 19$ is mainly due to the distance uncertainty cut we applied to the \citetalias{bailer-jones:21} distances. The majority ($53.3$ per cent) of stars in the Y1 subset are targeted as part of the {\tt BACKUP} programme, with the remaining stars distributed among the other target categories as follows: $20.5$ per cent {\tt MWS-BROAD}, $16.4$ per cent {\tt MWS-MAIN/FAINT-BLUE}, $0.4$ per cent {\tt MWS-MAIN/FAINT-RED}, and $9.4$ per cent {\tt SV1/2/3} targets \citep{cooper:23}.

Subsequently, we computed coordinates and velocities in a Galactocentric Cylindrical system $(R,\,\,\phi,\,\,Z)$ using \texttt{Astropy} \citep{astropy:13, astropy:18, astropy:22}. We adopted the following parameters: the distance from the Sun to the Galactic centre $R_{\odot} = 8.122$~kpc \citep{gravity:18}; the circular velocity at the Solar radius $v_c(R_{\odot}) = 229.0 \kms$ \citep{eilers:19}; the vertical height of the Sun from the Galactic plane $z_{\odot} = 20.8$~pc \citep{bennet_bovy:19}; and the solar motion in Galactic Cartesian coordinates $(v_X,\,\,v_Y,\,\,v_{Z})_{\odot} = (11.1, 12.24, 7.25) \kms$ \citep{schoenrich:10}. Since the Galactocentric frame in \texttt{Astropy} adopts a right-handed system, stars with prograde motion relative to the rotational direction of the Galactic disc have negative rotational velocities.

\begin{figure}
    \centering
    \includegraphics{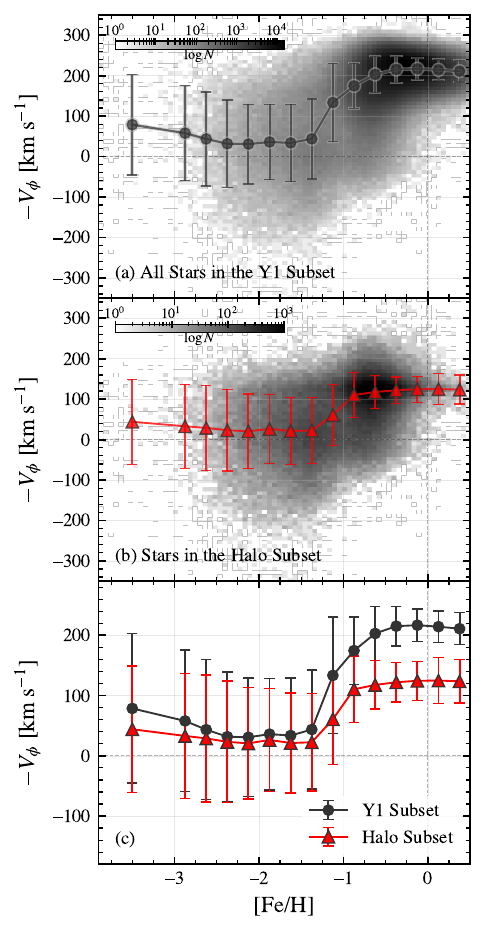}
    \caption{Rotational velocity in Galactocentric cylindrical coordinates as a function of stellar metallicity measured by the RVS pipeline. For visualisation purposes, we flip the sign of the rotational velocity, such that prograde motion corresponds to positive $\vphi$. The top (a) and middle (b) panels show the distributions of stars in the Y1 subset and the final halo subset, respectively. In each panel, filled points and error bars represent the median velocity of stars in each metallicity bin and standard deviation estimates from the difference between the 16th -- 84th percentiles. The bottom panel (c) compares the lines in panels (a) and (b). \label{fig:fig2}}
\end{figure}

Panel (a) in Figure~\ref{fig:fig2} presents the distribution of stars in the Y1 subset in rotational velocity ($-V_{\phi}$) -- metallicity ($\feh$) space. In this figure, we reverse the sign of rotational velocity to highlight velocity changes across the halo--disc transition. The majority of the Y1 subset shows disc characteristics, i.e., fast ($-\vphi \sim 220 \kms$) rotational velocity and higher metallicities ($\feh > -1.0$). A notable velocity spread begins to emerge around $\feh \sim -0.3$, indicating the presence of the Galactic thick disc and the {\it Splash} \citep{belokurov:20}. Stars with halo kinematics (i.e., lower $\vphi$) become prominent around $\feh \sim -1.0$. This disc--halo transition is more noticeable in the median $\vphi$ of all stars in each $\feh$ bin, shown by the black points. Error bars represent the 16th to 84th percentile range in $\vphi$ for each bin.

\subsection{Halo subset}\label{sec:halo_subset}
For the purpose of this study, we chose to use stellar orbital parameters to separate the halo population from the disc stars in the Y1 subset. For this, we adopted the \texttt{MilkyWayPotential} in \texttt{gala v1.5}\footnote{\url{https://gala.adrian.pw/en/latest/}} \citep{gala, gala_v15}, a four-component potential model comprised of a \citet{hernquist:90} bulge and nucleus, a Miyamoto-Nagai disc \citep{miyamoto-nagai:75}, and a Navarro-Frenk-White halo \citep[NFW;][]{nfw:97}, with the disc and bulge parameters fixed to those in \citep{bovy:15}. We implemented the \texttt{MilkyWayPotential} in \texttt{galpy v1.8.0}\footnote{\url{https://docs.galpy.org/en/v1.8.0/}} \citep{bovy:15} and calculate kinematic parameters, including total orbital energy ($\etot$), vertical component of angular momentum ($\lz$), and actions ($\jr$, $\jphi \equiv \lz$, $\jz$), making use of the Staeckel approximation \citep{binney:12, sanders:12} implemented in \texttt{galpy}. Additional orbital parameters, such as eccentricity ($e$), pericentre ($\rperi$), apocentre ($\rapo$), and maximum vertical height ($\zmax$), were determined by integrating orbits over a total integration time of 10 Gyr.

To obtain the halo subset, we excluded stars with disc kinematics using orbital circularity, $\epsilon$ \citep{abadi:03, grand:17, gomez:17}. The orbital circularity $\epsilon = L_z/L_{z, \mathrm{circ}}(E)$ conveys information about orbital eccentricity and the direction of orbital motion. $L_{z, \mathrm{circ}}(E)$ denotes the vertical angular momentum of an in-plane circular orbit at a given energy, $E$. Consequently, if a star with prograde motion follows a perfectly circular orbit, its circularity is $1$ (or $-1$ for a star with retrograde motion). In the halo subset, we only included stars with $\epsilon \leq 0.75$, resulting in a total of \num{138904} stars. Among those, we excluded $43$ stars having extremely large radial actions ($\jr > 10^4 \kpckms$\footnote{The $\jr$ cut is set at a value approximately an order of magnitude larger than the 95th percentile of the $\jr$ distribution ($\jr = 1.10 \times 10^3 \kpckms$) for the halo subset.}). These stars have high energy ($\etot > -0.35\times10^5 \kmkmss$), as well as high eccentricity ($e > 0.9$). About one-third of them ($15$ stars) seem to be unbound from the Galaxy, but this is due to their large distance error ($\sigma_{d_{\mathrm{helio}}}/d_{\mathrm{helio}} > 10$ per cent). As a result, \num{138861} stars remain in the halo subset. We note that all stars in the halo subset have projected rotational velocities, {\tt VSINI}, less than 1 $\kms$.

We investigated the fraction of metal-poor ($\feh < -1.0$) stars in the halo subset. For this, we computed the completeness and purity of low-metallicity stars in the halo subset, relative to the number of metal-poor stars in the Y1 subset. Before applying the circularity cut, there are \num{69013} stars with $\feh < -1.0$ in the Y1 subset. With \num{49344} metal-poor stars remaining in the halo subset, the completeness of the metal-poor population is $76.5$ per cent with a purity\footnote{The proportion of metal-poor stars (\num{69013}) among all stars (\num{138861}) in the halo subset.} of $35.5$ per cent. The apparent low purity is due to the contamination from the thick disc population. In the halo subset, about $38.2$ per cent of stars (\num{53171} stars) are low-mass dwarfs \citep[K and M dwarfs, $G - G_{\mathrm{RP}} > 0.56$ and $M_{G} > 5.55$;][]{pecaut:13} and about $80.7$ per cent of the sample (\num{112019} stars) are main-sequence stars ($4.0 < \logg < 5.0$).

The middle panel (b) in Figure~\ref{fig:fig2} shows the distributions of \num{138861} stars in the halo subset. The halo subset still shows the halo--disc transition due to a fraction of thick disc stars (red line), but it is less prominent than the Y1 subset (top panel). The overdensity around $\vphi \sim 100 \kms$ is also mainly due to the thick disc population with high circularity, close to the selection boundary ($\epsilon = 0.75$). 

The bottom panel (c) of Figure~\ref{fig:fig2} compares the median rotational velocity (filled points) of stars in each metallicity bin as a function of metallicity for the Y1 (black) and halo (red) subsets. Error bars represent velocity dispersion estimates from the difference between the 16th -- 84th percentiles. A small increase in $\vphi$ is observed in the Y1 subset at low metallicity ($\feh < -2.5$), mainly caused by M dwarfs in the disc that are affected by systematic effects from the stellar atmospheric models \citep{koposov:24}. This population is mostly removed after applying the circularity cut for the halo subset (red line). However, the disc--halo transition is still apparent in the halo subset.

\subsection{HDBSCAN*: Unsupervised clustering algorithm}
To search for stellar substructures in the halo subset, we used HDBSCAN* \citep[v0.8.33;][]{hdbscan, campello:13}, a publicly available unsupervised clustering algorithm optimised for finding over-dense regions in parameter spaces. It performs the data transformation, constructs the minimum spanning tree of the distance-weighted graph and a cluster hierarchy, merges the cluster hierarchy, and extracts the clusters from the condensed tree. More detailed information, including overviews and tutorials, is provided in the HDBSCAN* library documentation\footnote{\url{https://hdbscan.readthedocs.io/en/latest/how_hdbscan_works.html}}. As an ``unsupervised'' algorithm, HDBSCAN* does not require any prior knowledge or underlying assumptions about the data, which is the main advantage of using this method. 

HDBSCAN* is not the only unsupervised clustering algorithm in use, but it is one of the most accurate (high purity) and efficient (fast) \citep{hunt:21, brauer:22}. Due to its high performance and efficiency, the algorithm has been widely used in the astronomy community. One of its frequent applications is to identify substructures sharing similar kinematics and/or chemistry in chemo-dynamic spaces \citep{koppelman:19b, shank:22b, shank:22a, lovdal:22, shank:23, ou:23}. While those studies successfully detected and reported substructures, their typical approach involves using HDBSCAN* to break up the distribution in clustering spaces into small groups, followed by manual merging processes to identify larger associations. For this, it is necessary to define a small cluster size, less strict clustering conditions, and most importantly, a non-default clustering method called the `{\tt leaf}' method, in order to capture all small seed clusters of the decision tree. This common approach may certainly help in finding the substructures, but it seems to be somewhat distant from its intended use of unsupervised machine learning techniques for automated identification processes. In this work, therefore, we used HDBSCAN* with the default clustering method (`{\tt eom}' method) instead of the `{\tt leaf}' method. This allows the clustering algorithm to automatically join smaller clusters into the larger substructures to provide the final clustering results.

We ran the algorithm on two separate spaces: (1) the Integrals of Motion (IoM) space of energy, the vertical component of angular momentum, and the log-scaled radial and vertical components of the associated actions ($\etot$, $\lz$, $\log \jr$, $\log \jz$)\footnote{Creating a clustering space with three action components and total energy may introduce redundant information, as total energy is a function of the action components. However, including total energy does not significantly affect the clustering outcomes; the results are substantially consistent whether or not total energy is included in the clustering space. Given this, we have decided to keep total energy in IoM clustering space.}, and (2) the 3D Galactocentric cylindrical velocity space ($\vr$, $\vphi$, $\vz$). The IoM space is well-suited to detecting substructures that have been heavily phase-mixed. The purpose of applying a logarithmic scale to radial and vertical action components is to spread the data distribution, making clustering more efficient. Since the data distribution in both action components is more centrally concentrated with clear boundaries, it makes it difficult for the clustering algorithm to distinguish underlying substructures, especially in high-density areas. In contrast, the distribution of angular momentum is already well-spread and lacks the concentration seen in the action components, making additional scaling unnecessary. The velocity space, on the other hand, can be advantageous for detecting substructures from progenitors with low-velocity dispersion or those with high velocities found in velocity space, where there are not many stars in general. 

Before running the clustering algorithm, we transformed the IoM and velocity parameters using {\tt RobustScaler} from the {\tt scikit-learn} package \citep{scikit-learn}. {\tt RobustScaler} uses the median and interquartile range (25th -- 75th percentiles)\footnote{$X_{\mathrm{Scaled}} = (X - P_{50})/(P_{75} - P_{25})$, where X and P stand for a parameter value and percentile, respectively.} for scaling the data to a common range while ignoring outliers. The standardising process will make sure that outliers have less influence on the scaled data distribution and that each input parameter is treated equally by the clustering algorithm. 

Among the many tuning parameters for HDBSCAN*, the two most important are {\tt min\_cluster\_size} and {\tt min\_samples}. The {\tt min\_cluster\_size} sets the minimum number of data points that the algorithm considers as a cluster. The {\tt min\_samples} represents the number of neighbouring data points for a cluster's core point to be considered dense. A larger value of {\tt min\_samples}, hence, makes the algorithm more conservative in the final clustering as more points will be considered as background noise. Although fine-tuning these parameters is usually done empirically and intuitively (i.e., changing parameters until obtaining the best outcomes), we attempted to narrow down the ideal parameter range where we could obtain the best clustering results. To achieve this, we created $500$ mock datasets representing stellar populations in the Solar neighbourhood ($d < 10$~kpc) with substructure and ran the clustering algorithm with a range of parameters. We will provide more details in the next two sections.

\subsubsection{Mock data}\label{sec:mockdata}
To create mock data, we started with a simple assumption for how stellar populations of the local Galactic halo are distributed in position and velocity spaces. We disclaim here that our mock data are only used to understand how HDBSCAN* works and to decide which parameter ranges are useful for our analysis. We started with \num{100000} mock stars and only considered three main components in the local Galactic halo: $35$ per cent of thick disc stars, $50$ per cent of GSE stars, and $15$ per cent of the rest of the halo population, which is a combination of in-situ and accreted populations. The number ratio of the thick disc population is based on the number ratio of relatively metal-rich stars ($\feh > -0.7$) in the DESI halo subset. In the real local Galactic halo, the fraction of GSE stars has been estimated to be $15 - 25$ per cent \citep{lane:23}. We deliberately overrepresented GSE-like stars in our mock data in order to test how efficient HDBSCAN* is when it tries to detect a cluster with a few tens of members embedded in a dense background. We also note that a GSE-like structure is included not as an overdensity, but as part of a global distribution of local halo stars.

As we did not wish to include contributions from either the Galactic bulge or thin disc in the mock data, we used a uniform distribution in Galactic longitude, $l$. For Galactic latitude, $b$, we adopted a \emph{beta distribution} (with shape parameters $\alpha = \beta$) to generate a symmetric continuous distribution over the finite range $-90^{\circ} \le b \le 90^{\circ}$. For the halo population, we set the parameters $\alpha = \beta = 2$, which provides a gradual drop-off in density towards both Galactic poles; for the disc population, we set $\alpha = \beta = 5$, which has a more Gaussian-like profile and leads to a higher concentration near the Galactic plane.

\begin{figure*}
    \centering
    \includegraphics{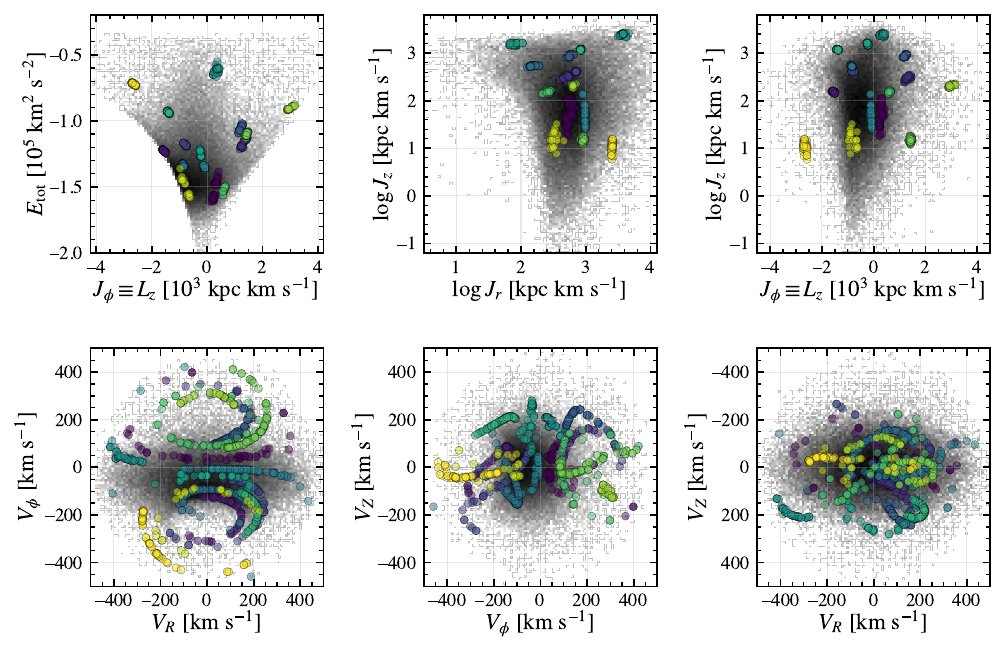}
    \caption{Kinematic distribution of mock stars in one of $500$ mock datasets. The top three panels show the distributions of background stars (grey 2D histogram) and stars in $15$ stellar streams (coloured points) in $\etot - \lz$ and log-scaled action spaces, while the bottom three panels show these in the Galactocentric cylindrical velocity spaces. \label{fig:fig3}}
\end{figure*}

A distance for each mock star is drawn randomly from the real halo subset, limited to distances smaller than $10$~kpc. We assumed that the velocity distribution in Galactocentric cylindrical coordinates ($\vr, \vphi, \vz$) for each stellar population follows a Gaussian distribution. We took velocity estimates for thick disc, GSE, and inner halo stars from \citet{belokurov:20} to define these distributions. After generating positions and velocities for the mock stars, we excluded stars with positive energies, high radial actions ($\jr \ge 10^4 \kpckms$) and large circularities ($\epsilon \ge 0.75$). Due to the circularity cut, the number of remaining stars in each mock dataset varies between $73,650$ and $74,350$. Figure~\ref{fig:fig3} presents one of the $500$ mock datasets in $\etot - \lz$, log-scaled actions, and Galactocentric cylindrical velocity spaces. The grey 2D histogram shows the distributions of background mock stars.

We added $15$ stellar streams to each mock dataset, each consisting of $10$ to $100$ stars. For this, we adopted the `particle-spray' stream generation method \citep{fardal:15} as implemented in the {\tt FardalStreamDF} class within the {\tt gala} package \citep{gala, gala_v15}. For each mock stream, the progenitor system is characterised by a Plummer potential featuring a randomly assigned stream mass ($m$) within a range of $10^4\, {\rm M}_{\odot}$ to $10^7\, {\rm M}_{\odot}$ and a core concentration factor ($b$) varying between $2$~pc and $7$~pc. Subsequently, we used {\tt MockStreamGenerator}, integrating it for the random time step ({\tt dt}) between $1$~Myr and $5$ Myr with $1,000$ steps. From this process, we obtained the orbital parameters of each mock stream star, and those with positive energies, high radial actions, and large circularity were excluded. The final members were then randomly selected to match the designated number of stars in each stream. All kinematic parameters of stars in the mock datasets were scaled using {\tt RobustScaler} in the same way as the kinematic parameters of the real data. 

Coloured circles in Figure~\ref{fig:fig3} represent the $15$ mock stellar streams. Each mock stream is tightly grouped in IoM space. The distributions in velocity spaces, on the other hand, appear to be in a thin line as mock stars in each stream have different phases around the orbit of each stream progenitor. These behaviours in different kinematic spaces directly affect the performance of HDBSCAN* (see \S\ref{sec:hyperparameter}).

\subsubsection{Hyperparameter tuning}\label{sec:hyperparameter}
To select the optimal values for the two main parameters of HDBSCAN*, {\tt min\_cluster\_size} and {\tt min\_samples}, we ran it on a grid of $156$ different parameter combinations. The grids of parameters are given below: 

\begin{itemize}
    \item {\tt min\_cluster\_size} = [{\tt 5, 7, 10, 15, 20, 25, 30, 40, 50, 70, 90, 100}] \\
    \item {\tt min\_samples} = [{\tt none, 5, 7, 10, 15, 20, 25, 30, 40, 50, 70, 90, 100}] \\
\end{itemize}

If {\tt min\_samples} remains unspecified ({\tt none}; default in HDBSCAN*), the parameter is set to be the same value as {\tt min\_cluster\_size}. Other parameters, such as {\tt metric} or {\tt cluster\_selection\_method}, were kept at their default values. 

Clustering with each parameter set was conducted in IoM space and Galactocentric cylindrical velocity space, respectively. HDBSCAN* provides an integer label for each star as an output, indicating the cluster to which it has been assigned. Stars belonging to the same cluster are given the same label, whereas stars that do not belong to any cluster are assigned a label of $-1$. Since we knew whether each of these stars was originally assigned to a mock stream (and if so, to which one), we compared their true labels with those assigned by HDBSCAN*. This allowed us to define `recovered' mock streams as clusters detected by HDBSCAN* that contain more than 50 per cent of the stars from any one of the 15 original mock streams.\footnote{It is possible for HDBSCAN* to split a single mock stream into multiple small clusters; in this case, we consider it as a single detection.} This procedure quantifies the purity and completeness of the clustering algorithm. The purity is defined as the ratio of recovered mock streams to the total number of clusters detected by HDBSCAN*, while the completeness is the ratio of recovered mock streams to the number of original mock streams ($15$). 

\begin{figure}
    \centering
    \includegraphics{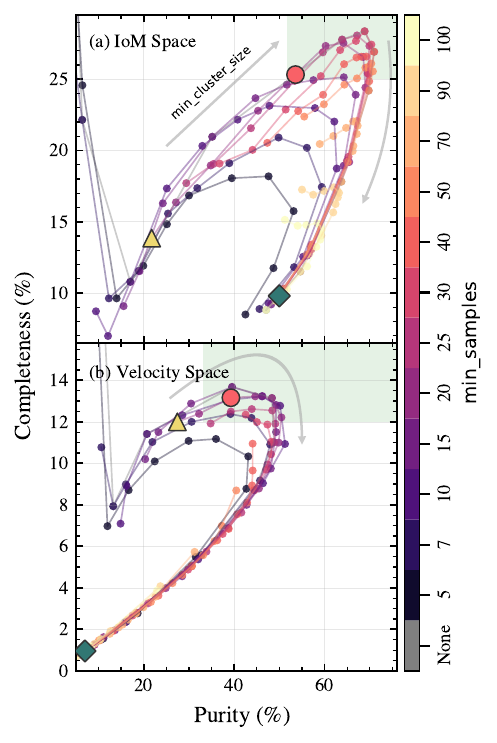}
    \caption{Changes in the completeness and purity of recovered mock streams with respect to different combinations of the parameters {\tt min\_cluster\_size} and {\tt min\_samples}. The top (a) and bottom (b) panels show clustering results in IoM and velocity spaces, respectively. Lines of different colours correspond to distinct values of {\tt min\_samples}, as shown in the colour bar. Along each line, points represent different choices of {\tt min\_cluster\_size}, increasing in the direction of the grey arrow. The red circles indicate the hyperparameter sets we adopt: $(20, 20)$ in IoM space and $(10, 20)$ in velocity space. The yellow triangles and green diamonds represent different examples of selecting smaller ($5$) or larger ($100$) {\tt min\_cluster\_size} while using the same {\tt min\_samples} of $20$. The green-shaded areas in both panels indicate the optimal high-completeness and high-purity parameter combinations. \label{fig:fig4}}
\end{figure}

Figure~\ref{fig:fig4} illustrates how completeness and purity vary over the grid of the two hyperparameters, {\tt min\_cluster\_size} and {\tt min\_samples}, in both IoM (top) and velocity (bottom) spaces. Each coloured line represents results from a different value of {\tt min\_samples} (see the colour bar in Figure~\ref{fig:fig4}); points along each line indicate increments of {\tt min\_cluster\_size} increasing in the direction shown by the grey arrows.

Generally, both completeness and purity peak for {\tt min\_cluster\_size} between $30$ and $40$; for larger values, completeness declines rapidly as {\tt min\_cluster\_size} increases. This is because larger values of {\tt min\_cluster\_size} will cause small mock streams to go undetected as they will not have a sufficient number of stars to be classified as a cluster by HDBSCAN*. The performance of HDBSCAN* is also affected by {\tt min\_samples}; optimal completeness and purity occur within the range of $10$ to $25$ for {\tt min\_samples}. Lower (higher) {\tt min\_samples} values provide suboptimal results; fewer (more) nearest neighbours result in less (more) restricted clustering, i.e., fewer (more) data points will be considered as background noise. 

In velocity space, the overall completeness and purity are lower compared to those obtained from clustering in IoM space. This decrease is likely because the mock streams are more dispersed in velocity space (see Figure~\ref{fig:fig3}). As a result, HDBSCAN* struggles to detect the overdensities corresponding to the mock streams. The clustering performance can improve by adjusting the hyperparameters, such as reducing {\tt min\_cluster\_size} or using {\tt cluster\_selection\_method == `leaf'}. These adjustments allow HDBSCAN* to split the same cluster into smaller, more detectable pieces.

We only considered the mock clustering results providing relatively high purity (IoM: $>50$ per cent, velocity: $>30$ per cent) and completeness (IoM: $>25$ per cent, velocity: $>12$ per cent) in both clustering spaces. These cases are found in the green-shaded area in Figure~\ref{fig:fig4}). Given this condition, the hyperparameter ranges we considered using for our analysis are the following (also see Table~\ref{tab:tab1}): 
\begin{itemize}
    \item {\tt min\_cluster\_size}: $15 - 50$ (IoM), $5 - 30$ (velocity)
    \item {\tt min\_samples}: $15 - 50$ (IoM), $7 - 30$ (velocity)
\end{itemize}

\begin{figure}
    \centering
    \includegraphics{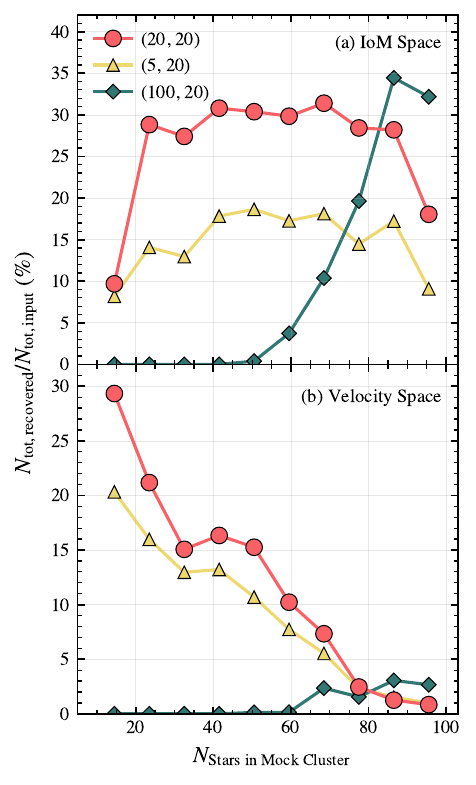}
    \caption{Completeness of mock clusters as a function of the number of stars in the mock cluster. A solid line with red-filled circles represents clustering results with adopted parameter combinations: {\tt min\_cluster\_size} $= 20$ and {\tt min\_samples} $= 20$ for the IoM space and {\tt min\_cluster\_size} $= 10$ and {\tt min\_samples} $= 20$ for the velocity space. We display two other scenarios using extreme values of {\tt min\_cluster\_size}, for comparison.: a solid line with yellow-filled triangles and a solid line with green-filled diamonds show the cases of using {\tt min\_cluster\_size} $= 5$ and $100$, respectively. \label{fig:fig5}}
\end{figure}

\begin{table}
    \centering
    \caption{HDBSCAN* hyperparameter combinations, providing high purity and completeness in both clustering spaces. \label{tab:tab1}}
    \resizebox{\columnwidth}{!}{%
    \begin{tabular}{cccc}
    \hline
    \multicolumn{2}{c}{IoM Space}  & \multicolumn{2}{c}{Velocity Space} \\
    {\tt min\_sample} & {\tt min\_cluster\_size} & {\tt min\_sample} & {\tt min\_cluster\_size} \\
    \hline
    \hline
        15 & 40 & 7 & 25, 30\\
        20 & 20, 25, 30, 40 & 10 & 20, 25, 30 \\
        25 & 20, 25, 30, 40, 50 & 15 & 10, 15, 20, 25 \\
        30 & 15, 20, 25, 30, 40, 50 & 20 & 7, 10, 15, 20, 25\\
        40 & 15, 20, 25, 30, 40 & 25 & 7, 10, 15\\
        50 & 15, 20, 25, 30, 40 & 30 &  5 \\
    \hline
    \end{tabular}}
\end{table}

In Figure~\ref{fig:fig5}, we show three examples presenting the completeness of recovered mock clusters as a function of the number of stars within the mock cluster, varying hyperparameter combinations. To compute the completeness, we first counted the total number of input and recovered clusters across $500$ simulations, respectively, consisting of the number of member stars, ranging from $10$ to $100$. These clusters were grouped into 10 bins based on the number of stars. The completeness was then calculated by dividing the total number of recovered clusters ($N_{\mathrm{tot, recovered}}$) by the total number of mock streams ($N_{\mathrm{tot, input}}$) in each bin\footnote{We also computed the mean completeness and purity for each cluster within each bin, based on the number of mock stars for each hyperparameter combination. The results are provided in two files (one for IoM and the other for velocity), available at the following link: \url{https://doi.org/10.5281/zenodo.15068958}.}. 

In IoM space, the completeness from our adopted parameter sets (red circles) becomes stable ($\sim30$ per cent) once mock streams consist of more than $20$ mock stars. Using a smaller minimum cluster size (yellow triangles) shows a similar trend, however, its completeness is significantly lower ($\sim15$ per cent) than for the adopted parameters. A larger minimum cluster size (green diamonds), on the other hand, yields higher completeness for mock clusters with more than $80$ stars ($> 25$ per cent) but entirely neglects smaller ones. Interestingly, the completeness with red circles and yellow triangles drops when the number of stars in the mock cluster is larger than $90$. This could be because larger mock streams are mistaken for part of the global distribution of mock stars. In velocity space, the red and yellow lines show comparable behaviours -- a declining trend in completeness with an increasing number of stars in the mock cluster --, but the red line presents slightly higher completeness for mock clusters with fewer than $60$ stars. The green line shows lower overall completeness (close to zero), possibly due to the dispersed distribution of mock streams in velocity spaces.

We tested different parameter combinations listed in Table~\ref{tab:tab1} and decided to use the hyperparameter pairs (shown as a red-filled circle in Figure~\ref{fig:fig4}) of ({\tt min\_cluster\_size}, {\tt min\_samples}) = $(20, 20)$ in IoM space and $(10, 20)$ in velocity space. Our choice prioritises combinations that maintain an equal number of detected clusters while avoiding any instances of splitting within these clusters. While our hyperparameter selection may appear to be a conservative approach, this procedure ensures both the consistency and stability of the identified clusters.

\section{Identification of Nearby Stellar Substructures}\label{sec:results}
From the previous discussion, we calibrated our best parameters for running HDBSCAN* and are now prepared to search for stellar substructures within the halo subset. Using the adopted hyperparameters, we ran HDBSCAN* on the halo subset, detecting several overdensities with similar dynamic properties in both IoM and velocity spaces. We classified the largest group detected in both clustering spaces as inner halo stars because they trace the overall distribution of the halo subset. More than $95$ per cent of stars in the halo subset are classified to belong to this group in both spaces, and about $50$ per cent of these stars are metal-rich ($\langle\feh\rangle \sim -0.8$), which once again confirms thick disc contamination in the halo subset. In addition to the group of inner halo stars, there are five (four) groups in IoM (velocity) space, which appear to be prominent overdensities with distinct chemical properties. In this section, we focus on these groups and discuss them in more detail.

\begin{figure*}
    \centering
    \includegraphics{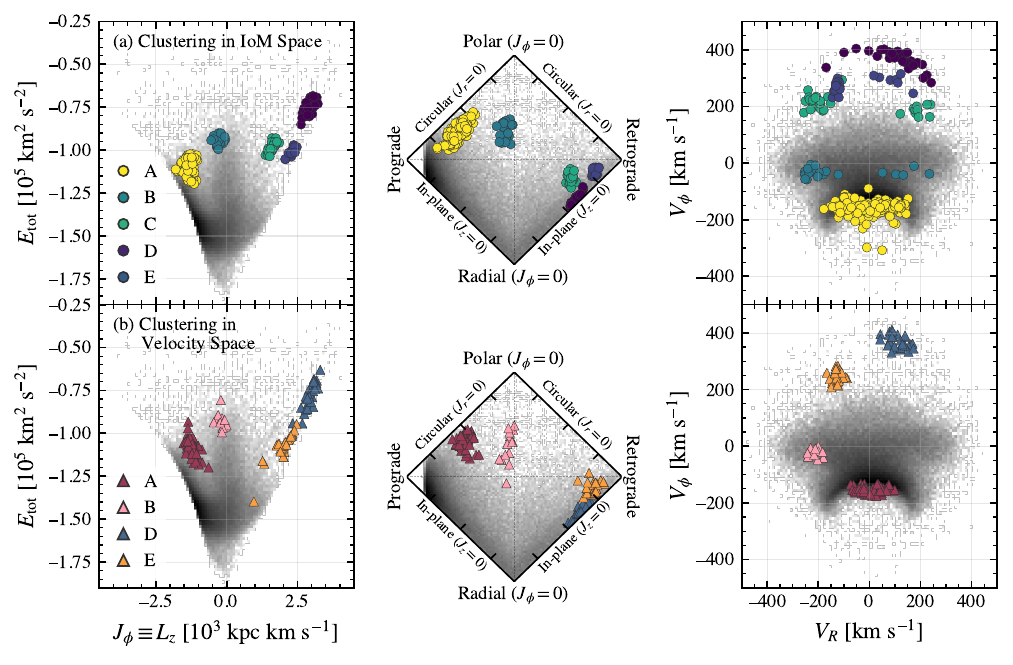}
    \caption{Distribution of the substructures identified by HDBSCAN* in $\etot - \lz$ space (left panels), the action diamond (middle panels), and $\vr - \vphi$ space (right panels). All stars in the halo subset are shown in the grey log-scaled 2D histogram, and coloured points represent clustered groups identified in the corresponding kinematic space. In the middle panels, the horizontal axis corresponds to $\jphi/J_{\mathrm{tot}}$, while the vertical axis indicates $(\jz - \jr)/J_{\mathrm{tot}}$, where $J_{\mathrm{tot}} = \jr+|\jphi|+\jz$.  \label{fig:fig6}}
\end{figure*}

Figure~\ref{fig:fig6} shows five (four) substructures identified by HDBSCAN* in IoM (velocity) space. The grey log-scaled 2D histogram represents the distribution of all stars in the halo subset. The top and bottom panels display clustering results in IoM and velocity space, respectively. The left, middle, and right panels show substructures in $\etot - \lz$ space, the normalised action diamond space \citep{vasiliev:19}, and $\vr - \vphi$ space, respectively. In the normalised action diamond space (middle panels), the horizontal and vertical axes correspond to $\jphi/J_{\mathrm{tot}}$ and $(\jz - \jr)/J_{\mathrm{tot}}$, respectively. The total action, $J_{\mathrm{tot}}$, is calculated from the sum of three action parameters, $\jr+|\jphi|+\jz$. 

All substructures detected in velocity space are associated with some of those detected in IoM space, sharing some number of stars in common. We merge clusters detected in both methods that have a large overlap in individual member stars, resulting in five clusters (see Table~\ref{tab:tab2}). The yellow group in the top panel and the dark red cluster shown in the bottom panel have the largest number of stars in common, so do dark green (top) - pink (bottom), dark purple (top) - dark blue (bottom), and dark blue (top) - orange (bottom). We merged these pairs to report the detection of five substructures in the DESI Y1 dataset. 

\begin{table}
    \centering
    \caption{The number of stars in common. \label{tab:tab2}}
    \begin{tabular}{rc}
    \hline
    Pair (IoM -- Velocity)$^a$ & Number of Stars \\
    \hline
    \hline
    Yellow -- Dark red & 57 \\
    Green -- Pink & 7 \\
    Dark purple -- Dark blue & 13 \\
    Dark blue -- Orange & 13 \\
    \hline
    \multicolumn{2}{l}{$^a$ Colours correspond to substructures in Figure~\ref{fig:fig6}.} \\
    \end{tabular}%
\end{table}

We initially name each structure as Clusters A (166 stars\footnote{The number of stars represents the total number of member stars after merging the groups from two searches.}; yellow in IoM + dark red in velocity), B (30 stars; dark green in IoM + pink in velocity), C (30 stars; light green), D (45 stars; dark purple in IoM + dark blue in velocity), and E (43 stars; dark blue in IoM + orange in velocity). Cluster A shows the most prograde motion among all the detected substructures, whereas Clusters C, D, and E show notable retrograde motion in both panels. Cluster B is characterised by its polar orbit and relatively higher energy. It is the smallest group reported in this work. Cluster C is solely detected in IoM space.

The stars in Clusters B, D, and E that are clustering in velocity space (i.e., the bottom panels in Figure~\ref{fig:fig6}) exhibit a larger spread in $\etot-\lz$ space and action diamond. Since some of these stars display significantly different orbital parameter distributions compared to other member stars in the same cluster, we consider them as outliers, coincidentally located in similar areas in velocity space. We exclude these stars lying outside the $95$ per cent confidence ellipse in IoM space, which is obtained by computing the median and covariance matrix of a set of the same four IoM parameters for clustering, assuming a Gaussian distribution. We iterate this process up to $15$ times, computing the new median and covariance matrix each time, until the number of selected stars stabilises. As a result, the final number of member stars remaining in each cluster is as follows: 121 in Cluster A, 24 in Cluster B, 27 in Cluster C, 42 in Cluster D, and 29 in Cluster E.

Table~\ref{tab:tab3} lists the five substructures we identify in the kinematic spaces. The table includes the total number of stars in each cluster along with the number of stars within the $95$ per cent confidence ellipse (given in brackets). It also provides the median and its uncertainty of each kinematic parameter, including $\etot$, $\lz$, $\jr$, $\jz$, $r_{\mathrm{peri}}$, and $r_{\mathrm{apo}}$, and the median metallicity and its dispersion. The rest of this paper refers to the value computed from the 16th -- 84th percentiles of the metallicity distribution of member stars found in the $95$ per cent confidence ellipse.

\begin{table*}
\centering
\caption{Five dynamic groups in the local Galactic halo detected in the DESI halo subset. \label{tab:tab3}}
\resizebox{\textwidth}{!}{%
\begin{tabular}{rccccccccccc}
\hline
     Cluster & ${N_{\mathrm{tot}}}^{a}$ & $\langle\etot\rangle$  & $\langle\etot\rangle_{95\%}$ & $(\langle\jr\rangle, \langle\jphi\rangle, \langle\jz\rangle)$ & $(\langle\jr\rangle, \langle\jphi\rangle, \langle\jz\rangle)_{95\%}$ & $\langle r_{\mathrm{peri}}\rangle$ & $\langle r_{\mathrm{peri}} \rangle_{95\%}$ & $\langle r_{\mathrm{apo}}\rangle$ & $\langle r_{\mathrm{apo}} \rangle_{95\%}$ & $\langle\feh\rangle^b$ & $\langle\feh\rangle_{95\%}^b$ \\ 
     & & $\sigma_{\langle\etot\rangle}$ & ${\sigma_{\langle\etot\rangle}}_{95\%}$ & $(\sigma_{\langle\jr\rangle}, \sigma_{\langle\jphi\rangle}, \sigma_{\langle\jz\rangle})$ & $(\sigma_{\langle\jr\rangle}, \sigma_{\langle\jphi\rangle}, \sigma_{\langle\jz\rangle})_{95\%}$ & $\sigma_{\langle r_{\mathrm{peri}} \rangle}$ & ${\sigma_{\langle r_{\mathrm{peri}}\rangle}}_{95\%}$ & $\sigma_{\langle r_{\mathrm{apo}} \rangle}$ & ${\sigma_{\langle r_{\mathrm{apo}}\rangle}}_{95\%}$ & & \\
 & & [$\kmkmss$] & [$\kmkmss$] & [$\kpckms$] & [$\kpckms$] & [kpc] & [kpc] & [kpc] & [kpc] &  &  \\
 \hline
 \hline
 A & $166 (121)$ & $-108826$ & $-106671$ & $(243.3, -1285.9, 1058.5)$ & $(238.6, -1283.3, 1139.8)$ & $7.62$ & $7.82$ & $17.16$ & $17.44$ & $-1.60_{-0.58}^{+0.41}$ & $-1.54_{-0.61}^{+0.34}$ \\ 
   &             & $4818$    & $4527$    & $(87.7, 114.8, 259.5)$     & $(74.8, 88.0, 225.7)$      & $0.92$ & $0.74$ & $2.15$  & $2.01$  &                              &                         \\
 B & $30 (24)$   & $-93933$  & $-93709$  & $(1017.1, -256.7, 2002.4)$ & $(1017.1, -278.4, 2002.4)$ & $5.90$ & $5.83$ & $27.28$ & $27.41$ & $-2.07_{-0.44}^{+0.25}$ & $-2.08_{-0.43}^{+0.19}$ \\
   &             & $3108$    & $3081$    & $(210.7, 112.4, 192.6)$    & $(192.6, 103.1, 145.3)$    & $0.41$ & $0.38$ & $2.44$ & $2.42$   &                              &                         \\
 C & $30 (27)$   & $-97155$  & $-97002$  & $(973.3, 1613.4, 449.9)$   & $(978.8, 1610.2, 447.5)$   & $5.10$ & $5.10$ & $25.21$ & $25.30$ & $-1.68_{-0.17}^{+0.35}$ & $-1.68_{-0.16}^{+0.39}$ \\
   &             & $2894$    & $2717$    & $(111.5, 100.8, 70.6)$     & $(99.7, 108.6, 61.7)$      & $0.34$ & $0.32$ & $1.73$ & $1.66$   &                              &                         \\
 D & $45 (42)$   & $-75925$  & $-75458$  & $(1907.2, 2918.0, 111.7)$  & $(1941.8, 2922.4, 114.3)$  & $7.46$ & $7.46$ & $43.86$ & $44.02$ & $-1.77_{-0.22}^{+0.33}$ & $-1.78_{-0.26}^{+0.31}$ \\
   &             & $7373$    & $5867$    & $(587.6, 154.5, 47.0)$     & $(527.2, 156.2, 44.9)$     & $0.49$ & $0.44$ & $9.12$ & $7.42$   &                              &                         \\
 E & $43 (29)$   & $-103144$ & $-101766$ & $(528.1, 2202.1, 203.0)$   & $(542.0, 2262.9, 193.2)$   & $6.90$ & $6.92$ & $20.52$ & $21.28$ & $-2.10_{-0.26}^{+0.66}$ & $-2.19_{-0.23}^{+0.37}$ \\
   &             & $4950$    & $4317$    & $(106.0, 235.1, 48.1)$     & $(108.6, 174.1, 45.4)$     & $0.49$ & $0.32$ & $2.53$ & $2.37$   &                              &                         \\
 \hline
\multicolumn{9}{l}{$^{a}$ The number of stars found in $95$ per cent confidence ellipse is shown in brackets.} \\
\multicolumn{9}{l}{$^{b}$ The median metallicity of stars is listed with an uncertainty computed from the 16th -- 84th percentiles of the metallicity distribution.} \\
\end{tabular}%
}
\end{table*}

\subsection{Validation with RVS Metallicities}\label{sec:validation}

\begin{figure}
    \centering
    \includegraphics{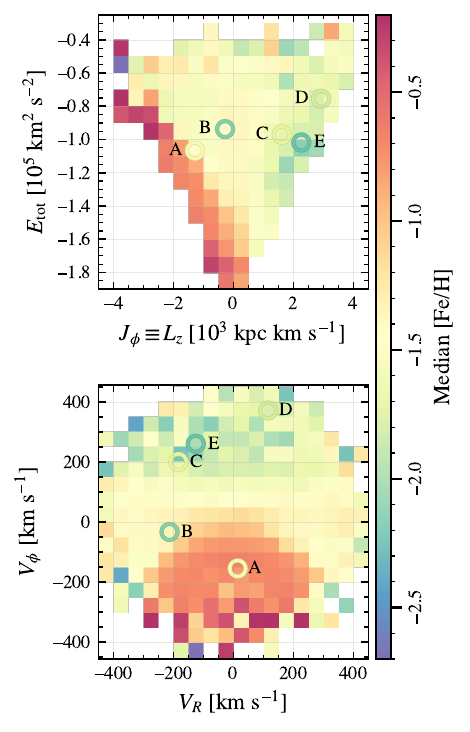}
    \caption{2D-histograms in $\etot - \lz$ (top) and the $\vphi - \vr$ space (bottom), colour-coded by median metallicity of each bin. Open circles show identified substructures in this study, with their loci representing median values of kinematic parameters. Each cluster is also colour-coded by its median metallicity for the visual metallicity comparison between detected clusters and inner halo stars. \label{fig:fig7}}
\end{figure}

We used DESI metallicities to validate the clustering results by comparing the metallicity distribution of each cluster with stars that lie in the same IoM space but are not included in the cluster (i.e., neighbouring inner halo stars). Figure~\ref{fig:fig7} shows 2D histograms of all stars in the halo subset, colour-coded by the median metallicity of each bin, in $\etot - \lz$ (top) and $\vr - \vphi$ (bottom). The five clusters identified in this study are shown as open circles, located at the median values of kinematic parameters ($\etot$, $\lz$, $\vr$, $\vphi$), and are also colour-coded by their median metallicities. This figure highlights how different the median metallicity of each cluster is compared to its neighbouring inner halo stars in the same IoM space. The clusters appear more prominent if they have significantly different median metallicities compared to neighbouring inner halo stars. As such, Clusters A and B appear more metal-poor compared to inner halo stars that do not belong to any dynamic groups, while the other three structures blend within the inner halo stars' metallicity distribution.

\begin{figure*}
    \centering
    \includegraphics{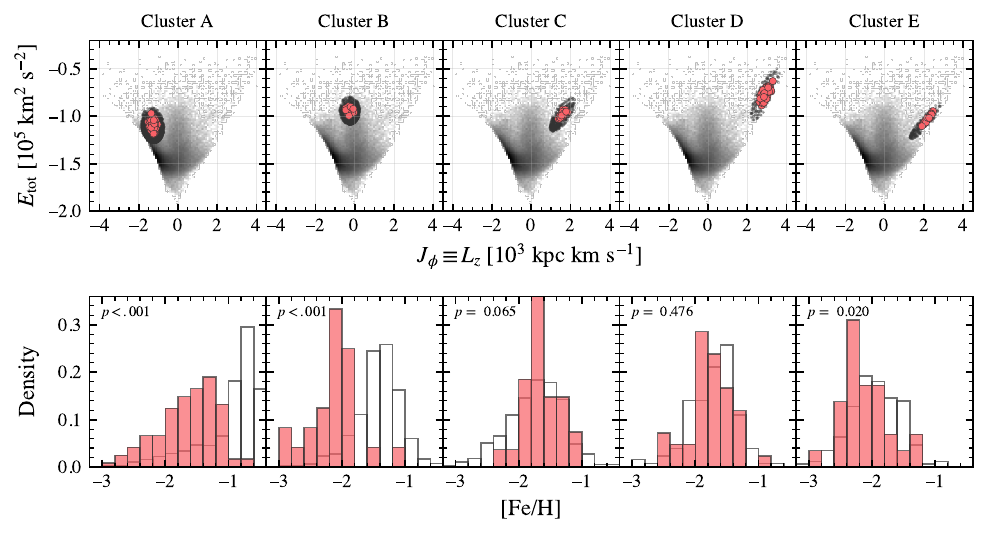}
    \caption{Comparison of metallicity distribution of each identified substructure with that of neighbouring background sources. The top panels display the distribution of all stars in the halo subset (greyscale 2D histogram), substructure member stars (red circles) selected from the $95$ per cent confidence ellipse, and neighbouring inner halo stars (black points) in $\etot - \lz$ space. The bottom panels show metallicity distributions of member stars selected from the confidence ellipse (red histogram) and neighbouring inner halo stars (black histogram). The metallicity distributions of Clusters A, B, and E are significantly different from those of inner halo stars, which is supported by low p-values. Clusters C and D, on the other hand, may appear chemically similar to the background stars. \label{fig:fig8}}
\end{figure*}

In Figure~\ref{fig:fig8}, we compare the metallicity distribution of member stars in each substructure with neighbouring inner halo stars with similar dynamic properties in order to check if the identified substructures are chemically distinct from the inner halo stars. The grey 2D histogram in the top panels represents the distribution of all stars in the halo subset. The neighbouring inner halo stars for each cluster (black points in the top panels) are selected in a confidence ellipse defined from the same covariance matrix of a set of IoM parameters but with a lower confidence level. The red-filled circles in the top panels represent the member stars located within the $95$ per cent confidence ellipse. The bottom panels present the metallicity distribution of the uncategorised (black histogram) and selected member stars (red histogram). The red histograms confirm that all identified substructures are consistently metal-poor. Cluster E ($\langle\feh\rangle = -2.19$) has the lowest median metallicity among our sample, followed by Cluster B ($\langle\feh\rangle = -2.08$). 

We ran the Kolmogorov–Smirnov test \citep[KS test;][]{kolmogoriv:33, smirnov:39} to compare the red-filled and black histograms and determine the similarity between these metallicity distributions; Clusters A, B, and E show the distinct metallicity distributions compared to their background counterparts ($p \le 0.020$). On the other hand, Clusters C ($p = 0.065$) and D ($p = 0.476$) appear to share a similar metallicity distribution to that of the background stars. Although there is a similarity shared by those groups and their background, we cannot simply rule out these clusters. Clusters C and D may be, in fact, part of the bigger substructures piling in retrograde, high-energy spaces. 

\subsection{Cluster stability}\label{sec:stability}
The clustering algorithm does not take measurement uncertainties into account. Depending on the size of those uncertainties, some clusters or member stars would have been excluded from the clustering. To understand the impact of uncertainties on the clustering, as well as gauge the significance of overdensities, we computed a `stability' rate of each cluster, which is defined as how many times each cluster has been detected when we draw a new phase-space measurement of each data point. 

For this, we assumed a Gaussian distribution of each astrometric and kinematic parameter (i.e., proper motions, distances, and radial velocities) with the reported measurement as the mean and its uncertainty as the standard deviation. With these new phase-space coordinates of each star, we repeated the same clustering procedure in IoM and velocity spaces. 

When we conducted a new clustering process on the newly defined datasets, it required caution to match the new results with the previous ones. Firstly, we defined an ellipsoid, which would contain $99$ per cent of the original member stars, using the covariance matrix of the original structure in each clustering space. If more than half of the stars of a newly detected cluster were found inside this ellipsoid, we considered it a recovered substructure. After iterating the process $200$ times, we obtained the stability rate by calculating the number ratio of recovered clusters to the total number of iterations. The whole process was performed separately in IoM space and velocity space.

\begin{table}
    \centering
    \caption{Stability rate of each cluster. \label{tab:tab4}}
    \begin{tabular}{rcc}
    \hline
             & \multicolumn{2}{c}{Stability} \\
     Cluster & IoM [\%]    &   Velocity [\%] \\
    \hline
    \hline
    A & $100$ & $99.5$  \\
    B & $23.5$ & $35.5$ \\
    C & $27.0$ & ---$^{a}$    \\
    D & $25.0$ & $99.5$ \\
    E & $40.5$ & $89.5$ \\
    \hline
    \multicolumn{3}{l}{$^{a}$ Cluster C is not identified in velocity space.} \\
    \end{tabular}%
\end{table}

In Table \ref{tab:tab4}, we list the stability of each substructure in IoM and velocity spaces, respectively. Most substructures discovered in velocity space have high stability ($> 89$ per cent), except for Cluster B ($35.5$ per cent), whereas most clusters found in IoM space show poor stability ($< 45$ per cent), except for Cluster A ($100$ per cent). Interestingly, the stability rates of Clusters C and D are low as they are found in IoM space with relatively low stellar number density compared to the other part of IoM space where lots of stars in the disc and GSE are found. It might be true that a relatively low background noise level (i.e., lower stellar number density) would help the clustering algorithm to easily find overdensities. However, since we perturbed the distribution of stars with uncertainties, clusters containing a small number of stars may have fewer stars remaining in the ellipsoid, which makes them less likely to be detectable. Furthermore, inner halo stars that were not originally included in the cluster may be found in the cluster with newly defined phase-space parameters because the parent substructure has a larger spread with uncertainties in the IoM space. This may explain why the original clustering did not find many stars in both clustering spaces that would actually belong to Clusters C and D. 

\section{Discussion}\label{sec:discussion}
\subsection{Association with known stellar substructures}\label{sec:association}
We compared the clusters identified in this study with reported substructures to find potential associations. The known substructures we discuss here include the Helmi streams \citep{helmi:99}, Candi10 \citep{myeong:18b}, Sequoia \citep{myeong:18a, myeong:18b}, Antaeus \citep{lovdal:22, ruiz-lara:22a, oria:22, ceccarelli:24}, and ED-2 \citep{dodd:23, lovdal:22, balbinot:23, balbinot:24}. Many of their member stars have also been targeted for spectroscopic follow-up observations \citep{koppelman:19a, aguado:21, matsuno:22a, matsuno:22b, oria:22, ceccarelli:24}. We collected information on the 6-D kinematics and metallicity from these studies and compared them with the member stars in each cluster. These substructures have the most similar chemodynamic properties to our clusters. Hence, we limited our search for the association to these five nearby known substructures.

\begin{figure*}
    \centering
    \includegraphics[]{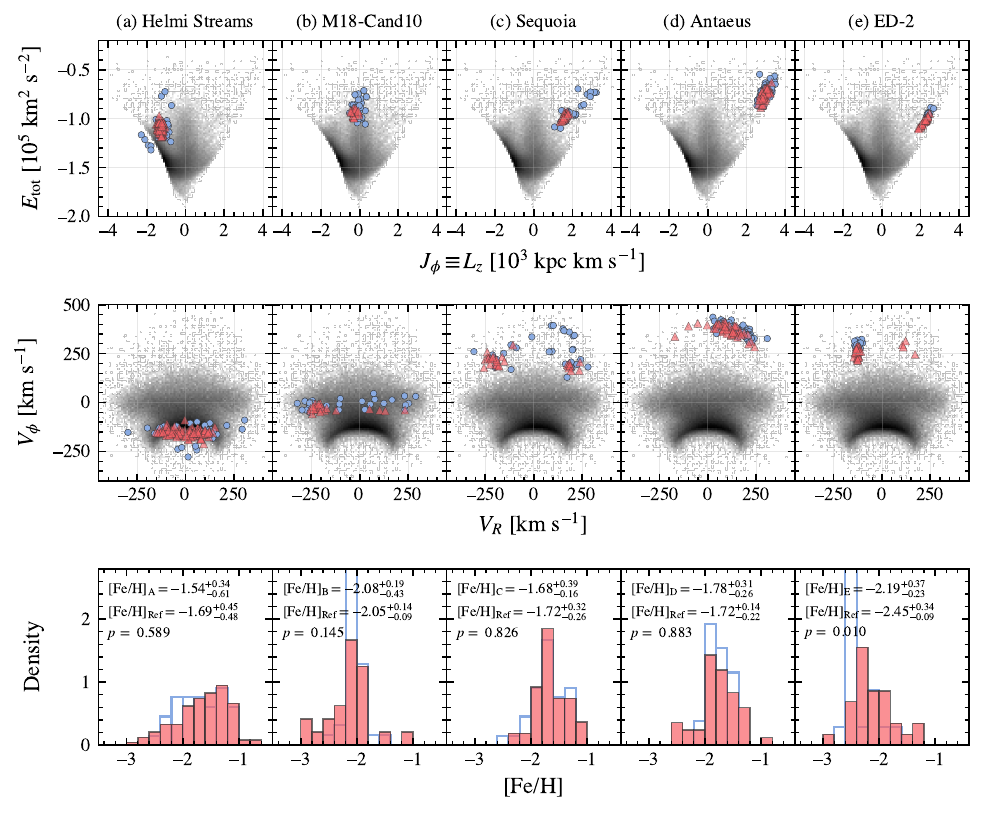}
    \caption{Comparison between our member stars and reported members from various studies in each cluster. The top and middle panels illustrate the distribution of member stars in $\etot - \lz$ and $\vphi-\vr$, respectively. Our member stars are shown as red triangles, while blue points represent known member stars from the literature, including the Helmi streams \citep{koppelman:19a}, M18-Candi10 \citep{myeong:18b}, Sequoia \citep{aguado:21, matsuno:22a, ceccarelli:24}, Antaeus \citep{oria:22, ceccarelli:24}, and ED-2 \citep{dodd:23, balbinot:23, balbinot:24, ceccarelli:24}. The bottom panels show the comparison between the metallicity distribution of our member stars (red histogram) in each detected cluster and that of reported members (blue histogram) in those clusters. The median metallicity and the dispersion of each substructure and the background are provided at the top of the bottom panels. The dispersion is measured from the difference between the 16th -- 84th percentiles of the distribution. The p-values from the KS test are also provided in the bottom row ($p$).  \label{fig:fig9}}
\end{figure*}

Figure~\ref{fig:fig9} presents the distribution of our clusters and the known substructures in $\etot-\lz$ (top row), $\vr - \vphi$ (middle row), and metallicity spaces. The grey 2D histograms in the top and middle rows display the distribution of the halo subset. Red points and histograms show DESI cluster member stars within the $95$ per cent confidence ellipse, while blue points and histograms represent member stars of known substructures from the literature, including the Helmi streams \citep[Panel a:][]{koppelman:19a}, M18-Candi10 \citep[Panel b:][]{myeong:18b}, Sequoia \citep[Panel c:][]{aguado:21, matsuno:22a, ceccarelli:24}, Antaeus \citep[Panel d:][]{oria:22, ceccarelli:24}, and ED-2 \citep[Panel e:][]{dodd:23, balbinot:23, balbinot:24, ceccarelli:24}. The visual representations in different kinematic spaces offer a clear view of the similarities between our member stars and those reported in the literature. 

We briefly discuss each substructure below, including a summary of the previous research. Tables~\ref{tab:tab5} and \ref{tab:tab6} provide column description and a sample table of the list of member stars in each substructure, including their DESI {\tt TARGET\_ID}, {\it Gaia} DR3 {\tt source\_id}, radial velocities, and stellar parameters obtained from DESI. The full table of \num{314} member stars (Table~\ref{tab:tab6}) is available online in a machine-readable format\footnote{\url{https://doi.org/10.5281/zenodo.15068958}}.

\begin{table}
\caption{Column names and descriptions of the list of member stars in five dynamic groups in the halo subset.}
\centering
\label{tab:tab5}
\resizebox{\columnwidth}{!}{%
\begin{tabular}{cl}
\hline
Header & Description \\
\hline
\hline
 {Cluster} & Cluster label (A, B, C, D, and E) \\
 {\tt TARGETID} & DESI MWS Y1 unique source identifier \\
 {\tt source\_id} & {\it Gaia} DR3 unique source identifier \\
 R.A. & Right ascension in J2016.0 from {\it Gaia} DR3 [deg] \\
 Decl. & Declination in J2016.0 from {\it Gaia} DR3 [deg]  \\
 $\mu_{\alpha}^{*}$ & Proper motion in right ascension direction from {\it Gaia} DR3 [mas yr$^{-1}$] \\
 $\mu_{\delta}$ & Proper motion in declination direction from {\it Gaia} DR3 [mas yr$^{-1}$] \\
 $d$ & Distance [kpc] \\
 Distance Ref. & Distance references (chosen from either {\it Gaia} DR3 parallax or \citetalias{bailer-jones:21}) \\
 {$V_{\mathrm{rad}}$} & Radial velocity from DESI RVS pipeline [$\kms$]\\
 {$\sigma_{V_{\mathrm{rad}}}$} & Uncertainty in radial velocity from DESI RVS pipeline [$\kms$]\\
 {$\feh$} & Metallicity estimated from DESI RVS pipeline [dex] \\
 {$\sigma_{\feh}$} & Metallicity uncertainty estimated from DESI RVS pipeline [dex] \\
 {$\etot$} & Total orbital energy [$\kmkmss$] \\
 {$\jr$} & Radial orbital action [$\kpckms$] \\
 {$\jphi \equiv \lz$} & Azimuthal orbital action (angular momentum in vertical direction) [$\kpckms$] \\
 {$\jz$} & Vertical orbital action [$\kpckms$] \\
 {$\vr$} & Radial velocity in Galactocentric cylindrical coordinates [$\kms$] \\
 {$\vphi$} & Azimuthal velocity in Galactocentric cylindrical coordinates [$\kms$] \\
 {$\vz$} & Vertical velocity in Galactocentric cylindrical coordinates [$\kms$] \\
 {Footnote} & Footnote indicators (a, b)\\
\hline
\end{tabular}%
}
\end{table}

\begin{table*}
\caption{A list of \num{314} member stars in five dynamic groups in the halo subset compiled with {\it Gaia} DR3.\label{tab:tab6}}
\resizebox{\textwidth}{!}{%
\centering
\begin{tabular}{rcccccccccc}
\hline
 {Cluster} & {DESI {\tt TARGETID}} & {{\it Gaia} DR3 {\tt source\_id}} & {...} & {$d$} & {Distance Ref.} & {$V_{\mathrm{rad}}$} & {$\sigma_{V_{\mathrm{rad}}}$} & {$\feh$} & {$\sigma_{\feh}$} & {...} \\
 {} & {} & {} & {...} & [kpc] & {} & [{$\kms$}] & [$\kms$] &  {} & {} & {...} \\
\hline
\hline
  A & 2305843036791244973 & 3701636431849634176 & {...} & 2.4 & \citetalias{bailer-jones:21} & 247.56 & 1.44 & $-1.80$ & 0.08 & {...} \\
  A & 2305843017640056412 & 1131202305964045696 & {...} & 0.4 & Parallax & $-191.33$ & 1.21 & $-1.47$ & 0.03 & {...} \\
  A & 2305843021716916001 & 1678266515386942464 & {...} & 1.0 & Parallax & $-279.06$ & 0.62 & $-1.46$ & 0.04 & {...} \\
  A & 2305843018504081012 & 1247487720868852864 & {...} & 2.8 & \citetalias{bailer-jones:21} & 185.63 & 0.98 & $-1.13$ & 0.05 & {...} \\
  {...} & {...} & {...} & {...} & {...} & {...} & {...} & {...} & {...} & {...} & {...} \\
\hline
\multicolumn{11}{l}{Note. See also Table~\ref{tab:tab5} for column description. A full machine-readable table is available online: \url{https://doi.org/10.5281/zenodo.15068958}} \\
\multicolumn{11}{l}{a. Stars found in $95$ per cent confidence ellipse.} \\
\multicolumn{11}{l}{b. Cluster E member stars with positive radial velocities, $\vr > 0\,\,\kms$.} \\
\end{tabular}%
}
\end{table*}
  
\subsubsection{Cluster A: Helmi streams}
Cluster A is the largest dynamical group identified in this study. It has been detected in both clustering spaces, and it consists of $166$ ($122$) stars from the initial ($95$ per cent confidence ellipse) selection. It shares similar kinematic properties with the {\it Helmi streams} \citep{helmi:99}, which is the first substructure found in the Galactic inner halo by using IoM space. \citet{helmi:99} reported a group of metal-poor stars in the Solar neighbourhood, clumping in IoM space ($\jz$ and $J_{\perp}$), and they propose that the group originated from a disrupted accreted dwarf galaxy similar to the Fornax spheroidal galaxy. \citet{koppelman:19a} and \citet{ruiz-lala:22b} provided a detailed analysis of the Helmi streams using the second and early third data releases from {\it Gaia} \citep{gaia:18, gaia:21}, implying a stellar mass of $M \sim 10^8\,\mathrm{M}_{\odot}$ and its accretion time (8 Gyr ago). Spectroscopic follow-up studies \citep{koppelman:19a, limberg:21, matsuno:22b} revealed that the metallicity distribution of stars in the Helmi streams spans between $-2.5$ dex and $-1.5$ dex with a mean metallicity of $-1.5$ dex.

\citet{koppelman:19a} reported two lists of the Helmi streams member stars: 40 core and \num{523} Selection B members\footnote{Cluster A member stars in $95$ per cent confidence ellipse shares five stars with the Selection B Helmi streams members, and the {\it Gaia} DR3 {\tt source\_id}s for those stars are the following: 3902635335025505408, 3842793349531132928, 841170654888167808, 1527475951701753984, and 1512783143459215744.}. We used a list of core members from \citet{koppelman:19a} and computed their kinematic properties as described in \S\ref{sec:halo_subset} (Blue circles in the left panels in Figure~\ref{fig:fig9}). Since we did not apply the same circularity cut to the literature data as we did to define our halo subset, a few Helmi streams member stars rotating faster than thick disc stars remain in this analysis and are shown to be outside the distribution of all stars in the DESI halo subset. Cluster A and the group of core members from \citet{koppelman:19a} both show prograde motion and relatively circular orbits compared to other substructures we detect, while they appear to occupy a range of velocities. In velocity space (left middle panel in Figure~\ref{fig:fig9}), Cluster A appears to fill the velocity gap in $\vphi$ between the two groups of the Helmi streams, i.e., stars with faster $\vphi$ and those with slower $\vphi$.

The bottom-left panel of Figure~\ref{fig:fig9} shows the metallicity distribution of our member stars (red histogram) and stars in the Helmi streams with metallicities reported from previous studies (blue histogram), including \citet{limberg:21} and \citet{matsuno:22b}. The median metallicity and its 16th -- 84th percentile dispersion of \num{121} Cluster A stars are $\feh_{\mathrm{A}} = -1.54^{+0.34}_{-0.61}$ and $\feh_{\mathrm{ref}} = -1.69^{+0.45}_{-0.48}$ for the $33$ Helmi streams stars. The KS test confirms that the distributions are consistent ($p = 0.970$), and their median metallicities agree within each other's $1\sigma$ intervals. Since we compared DESI metallicities without calibrating with different studies, this difference could arise from the significant systematic difference between spectroscopic surveys.

\subsubsection{Cluster B: M18-Cand10 / MMH-1}\label{sec:cluster_b}
Cluster B is detected in both IoM and velocity spaces. It consists of $30$ stars that are initially selected from both spaces, out of which $24$ stars are found in the $95$ per cent confidence ellipse. Cluster B seems to follow a polar orbit, showing slightly prograde motion. Its kinematic properties appear similar to those of GSE stars with higher energy based on its distribution on the $\etot - \lz$ space, which may suggest that it could have been considered part of the GSE. However, its metallicity distribution ($\langle\feh\rangle = -2.08$) is noticeably different compared to that of stars sharing similar $\etot$ and $\lz$ ($\langle\feh\rangle \sim -1.39$; see Figure~\ref{fig:fig8}). One possible explanation is that Cluster B is a high-energy debris of GSE, consisting of stars located at the outskirts of the GSE that were stripped first during the merger \citep{koppelman:20}. Assuming a metallicity gradient in GSE \citep{khoperskov:23b}, it is possible that high-energy GSE stars are more metal-poor than other GSE stars that were accreted later.

Considering its kinematic characteristics, it appears to be associated {\it Candidate 10 (M18-Cand10)} reported in \citet{myeong:18b}. They reported a few groups with polar orbits and low metallicities using the SDSS--{\it Gaia} DR2 catalogue. Among those, M18-Cand10 shows the most similar kinematic properties and metallicity to Cluster B. The actions and metallicity information of M18-Cand10 reported in \citet{myeong:18b} show relatively high vertical action ($\langle\jz\rangle = 2345.9 \kpckms$, $\sigma_{\jz} = 329.2 \kpckms$) with slightly prograde polar motion ($\langle\jphi\rangle = -94.2 \kpckms$, $\sigma_{\jphi} = 144.5 \kpckms$)\footnote{\citet{myeong:18b} adopted a left-handed system, which lets a star with prograde motion relative to the rotational direction of the Galactic disc shows positive angular momentum. We flip the sign to keep consistency in this paper. We also note that \citet{myeong:18b} used the Galactic potential from \citet{mcmillan:17}.}, and low mean metallicity ($\langle\feh\rangle = -2.01$, $\sigma_{\feh} = 0.18$), which are comparable with Cluster B's properties (e.g., $(\langle\jphi\rangle,\,\,\sigma_{\langle\jphi\rangle},\,\,\langle\jz\rangle,\,\,\sigma_{\langle\jz\rangle})_{\mathrm{95\%, B}} = (-278.4,\,\,103.1,\,\,2002.4,\,\,145.3) \kpckms$, and $\langle\feh\rangle_{\mathrm{95\%, B}} = -2.08$). 

To pin down the potential association, we compared the kinematic and metallicity distributions of both groups. We adopted the distances and radial velocities from \citet{myeong:18b}\footnote{Data was obtained from private correspondence.} and computed the orbit properties of the stars in M18-Cand10, assuming the same conditions that we have described above (see \S~\ref{sec:halo_subset}). The sky positions and proper motions of M18-Cand10 members were updated by cross-matching with {\it Gaia} DR3. We crossmatched M18-Cand10 member stars to the Sloan Extension for Galactic Understanding and Exploration (SEGUE) survey catalogue \citep{yanny:09} to obtain their metallicities measured from the SEGUE Stellar Parameter Pipeline \citep[SSPP;][]{lee:08a, lee:08b, allende_prieto:08}.

The second column of Figure~\ref{fig:fig9} shows the comparison between M18-Cand10 and Cluster B\footnote{Member stars of Cluster B found in $95$ per cent confidence ellipse share two stars with M18-Cand10. {\it Gaia} DR3 {\tt source\_id}s of those two stars are the following: 3676209469543208576 and 1482427860799814016.}. The top and middle panels show the distribution of both groups (M18-Cand10: blue-filled circles, Cluster B: red-filled triangles) in kinematic space, while the bottom panel compares the metallicity distribution of both groups (M18-Cand10: black histogram, Cluster B: red-filled histogram). Both groups appear to share very similar chemodynamic properties: the distribution of Cluster B member stars overlays closely with that of M18-Cand10, indicating their kinematic resemblance. M18-Cand10 occupies a wide range of velocities, and Cluster B also appears to undergo a phase-mixing process, given its distribution in the second middle panel of Figure~\ref{fig:fig9}. The average metallicity of Cluster B ($\langle\feh\rangle = -2.08$), calculated from stars within $95$ per cent confidence ellipse, is also in good agreement with that of M18-Cand10 ($\langle\feh\rangle = -2.05$). This similarity is supported by the KS test ($p = 0.145$).

\begin{figure*}
    \centering
    \includegraphics{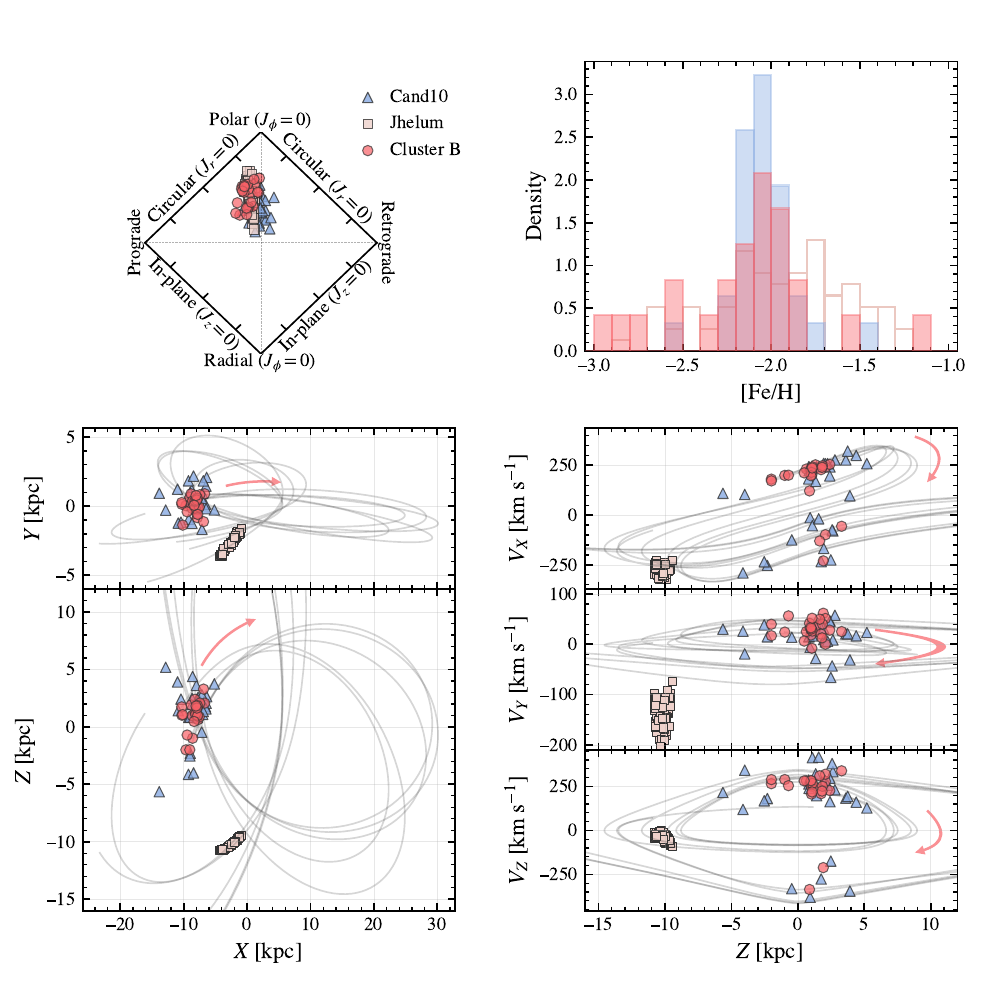}
    \caption{Comparison of the characteristics between Cluster B, M18-Cand10, and Jhelum. Member stars of M18-Cand10 are from \citet{myeong:18b}, and Jhelum member stars are from \citet{li:19} and \citet{li:22}. {\it Top-left panel:} Three dynamic groups in an action diamond. Stars in all three dynamic groups have polar orbits. Jhelum (beige squares) and Cluster B (red circles) have slightly more prograde motion compared to M18-Cand10 (blue triangles). {\it Top-right panel:} Metallicity distributions of three dynamic groups. As shown in Figure~\ref{fig:fig9}, the metallicity distribution of Cluster B (red histogram) is very similar to that of M18-Cand10 (blue histogram), whereas Jhelum (beige histogram) has a broader metallicity distribution extended towards the more metal-rich ($\feh > -1.8$) regime. {\it Bottom panels:} Stars in Cluster B and M18-Cand10 in position and velocity spaces ($X,\,\,Y,\,\,Z,\,\,V_X,\,\,V_Y,\,\,V_Z$) with five representative trajectories (grey lines) of Cluster B forward integrated over $900$ Myr from now. The red arrows show the direction of Cluster B's orbital motions. \label{fig:fig10}}
\end{figure*}

Figure~\ref{fig:fig10} displays the kinematic distributions of M18-Cand10 and Cluster B stars with blue-filled circles and red-filled triangles, respectively. Both groups appear to show comparable action (top-left panel in Figure~\ref{fig:fig10}) and spatial distribution (bottom-left panels in Figure~\ref{fig:fig10}). It seems that M18-Cand10 shows more spread in velocity space (bottom-right panel in Figure~\ref{fig:fig10}), which may indicate that M18-Cand10 experiences more phase-mixing. Despite this difference, the similarity in their orbital characteristics and metallicity distributions suggests an association between the two groups.

{\it MMH-1} \citep{mikkola:23} is another possible dynamic group that could be associated with Cluster B. It is a recently reported feature found in the probability distribution of velocities of stars from {\it Gaia} DR3, specifically in the blue main sequence, as represented in spherical coordinates \citep{mikkola:23}. We compared the velocity distribution of Cluster B in the Galactocentric spherical coordinates to the probability distribution in \citet{mikkola:23}\footnote{The probability distribution was obtained from private correspondence.}. Cluster B shows median velocities in Galactocentric spherical coordinates of $\langle(\vr, \vphi, V_{\theta})\rangle_{\mathrm{95\%, B}} = (-213.1, -33.3, 266.5)\kms$ with velocity dispersions of $\sigma_{(\vr, \vphi, V_{\theta})_{\mathrm{95\%, B}}} = (87.3, 12.1, 36.8)\kms$, and its distribution is close (within $1\sigma$-level) to the feature where MMH-1 is defined, $\langle(\vr, \vphi, V_{\theta})\rangle_{\mathrm{MMH-1}} = (\pm225, -25, 325)\kms$\footnote{\citet{mikkola:23} adopted a left-handed system with respect to the rotational direction of the Galactic disc. We flipped the sign of $\vphi$ from \citet{mikkola:23} to make it consistent with this work.}. This suggests that MMH-1 is likely to be associated with Cluster B as well as M18-Cand10.

Interestingly, Jhelum stellar stream \citep{shipp:18, malhan:18b} is located in the same area in both $\etot-\lz$ and action diamond spaces where Cluster B and M18-Cand10 are found. To compare its kinematic properties with Cluster B, we obtain astrometric parameters (position, proper motions), distance estimates, radial velocities, and metallicities of $95$ Jhelum members from the latest data of the Southern Stellar Stream Spectroscopic Survey \citep[S5;][]{li:19, li:22}. Their orbital parameters were computed in the same way as in \S~\ref{sec:y1subset} and \S~\ref{sec:halo_subset} throughout the paper that we used for DESI stars.

The distribution of Jhelum stars in the top-left panel of Figure~\ref{fig:fig10} confirms the similarity in orbital parameters between Jhelum and Cluster B. Despite this, the metallicity distribution of Jhelum (beige-unfilled histogram in Figure~\ref{fig:fig10}) obtained from the S5 survey is not entirely similar to that of Cluster B ($p = 0.033$), even though the median metallicity ($-1.90$ dex) of Jhelum seems to be comparable to that of Cluster B. The discrepancy in the metallicity distributions seems to occur due to relatively metal-rich stars ($\feh > -1.8$).

We integrated the orbits of Cluster B stars to confirm their potential association with Jhelum. The bottom panels of Figure~\ref{fig:fig10} show the forward integrated orbits (black solid lines) of five randomly selected Cluster B stars as well as the current location of M18-Cand10 (blue-filled triangles) and Jhelum (beige-filled squares) in the position and velocity planes in Galactocentric Cartesian coordinates, $XYZ$. The red arrows indicate the directions of Cluster B, and the future trajectories of five Cluster B stars are integrated over the next $900$ Myr. The backwards integrated orbits over the same amount of time do not show any sign of association, therefore, we choose not to show those orbits for clarity.

Cluster B and M18-Cand10 share very similar distributions in position and velocity spaces, whereas the orbits computed in one orbital cycle timescale ($t_{\mathrm{integrate}} \sim 380$ Myr) of Cluster B do not converge in the current position of Jhelum. While the orbits integrated for a much longer timescale ($t_{\mathrm{integrate}} = 900$ Myr) appear to intersect at Jhelum, velocities in $Y$ ($V_Y$) of Cluster B still show huge discrepancy, as shown in the middle panel of the bottom-right panel in Figure~\ref{fig:fig10}. Given the same amount of time, not only will Cluster B pass the same location where Jhelum is currently located, but it will also have very similar velocities in $V_X$ and $V_Z$ (see the bottom-right panels in Figure~\ref{fig:fig10}). While other parameters are in good agreement, the discrepancy regarding matching trajectories arises in $V_Y$. Jhelum's median values of $V_Y$ seem to differ from the future trajectories of Cluster B by approximately a $3\sigma$ level. 

Due to this discrepancy, we rule out the possibility of Cluster B being associated with Jhelum. While all the other kinematic parameters are in good agreement, the difference in $V_Y$ is significant, which implies these two groups are not associated with each other. Given the similarities of both dynamic groups in action diamond and metallicity distribution, however, another possibility is that they originated from the same progenitor but then separated from each other due to the encounter with a massive object, such as the Sagittarius dwarf galaxy \citep{woudenberg:23}. 

\subsubsection{Clusters C and D: Sequoia and Antaeus\protect\footnotemark}
\footnotetext{While we identified and characterised these two are different dynamic groups, we discuss Clusters C and D together in this section, given the customary selection of Sequoia's member stars.}
Cluster C is only detected in IoM space and consists of $30$ stars, showing retrograde motion from the initial selection, making it one of the smallest kinematic groups identified in this study, along with Cluster B. The median metallicity of stars found in $95$ per cent confidence ellipse is $-1.68$, which is similar to that of background stars ($\langle\feh\rangle = -1.72$), however, the overall metallicity distribution (bottom-middle panel in Figure~\ref{fig:fig8}) seems to have a smaller dispersion ($\sigma_{\feh} = 0.28$) than that of background stars ($\sigma_{\feh} = 0.44$). It appears to share similar energy and angular momentum with {\it Sequoia}, which was first identified by \citet{myeong:18a, myeong:18b} using the {\it Gaia} DR1 \citep{gaia:16} compiled with SDSS-DR9.

The third column of Figure~\ref{fig:fig9} presents the distribution of Cluster C (red-filled triangles) and Sequoia \citep[blue-filled circles;][]{aguado:21, matsuno:22a, ceccarelli:24}. The distribution of Sequoia stars from the literature spans a wide range of total energy and velocities (top and middle panels), as both spectroscopic surveys applied a simple selection box in high-energy, retrograde $\etot - \lz$ space \citep[See also][]{aguado:21, matsuno:22a, koppelman:19b}. Interestingly, Sequoia appears to consist of two groups with slightly different energies (i.e., high- and low-energy groups), and Cluster C overlaps only with the low-energy group.

The high-energy group of Sequoia seems to share characteristics more similar to Cluster D, another retrograde group detected in our study. It consists of $45$ stars from the initial selection in both kinematic spaces. It has the highest energy ($\langle\etot\rangle_{95\%} = -75925 \kmkmss$) among all the substructures identified in this study. While it has very retrograde motion ($\langle\jphi\rangle_{95\%} = 2918.0 \kpckms$), it exhibits very low vertical action ($\langle\jz\rangle_{95\%} = 111.7 \kpckms$), which makes it different from Sequoia.

Given the original selection for Sequoia from \citet{myeong:18a, myeong:18b} or \citet{naidu:20}, Cluster D might be associated with Sequoia as well. However, the latest development in selecting Sequoia member stars has suggested different possibility; \citet{ruiz-lara:22a} and \citet{oria:22} independently reported a very retrograde dynamic group with low $\jz$, {\it \#64} or {\it Antaeus} \citep[hereafter Antaeus; see also][]{lovdal:22, ceccarelli:24}. Its action properties (i.e., $\langle(\jr, \jphi, \jz)\rangle = (1761, 2990, 39)\kpckms$)\footnote{\citet{oria:22} adopted the Galactic potential from \citet{mcmillan:17}.} are very similar to those of Cluster D ($\langle(\jr, \jphi, \jz)\rangle_{\mathrm{95\%, D}} = (1941.8, 2922.4, 114.3)\kpckms$; also see the top and middle panels in Figure~\ref{fig:fig9}), as well as the low $\jz$. We compared the metallicity distribution of $13$ Antaeus stars found in common in both \citet{oria:22} and \citet{ceccarelli:24} with that of Cluster D stars in the bottom panel of Figure~\ref{fig:fig9}. The mean metallicity of Antaeus stars is $-1.72$, which is consistent with what we have measured for Cluster D ($\langle\feh\rangle = -1.78$ with a dispersion of $0.29$ dex). Therefore, we conclude that Cluster D is likely to be associated with Antaeus.

\citet{ceccarelli:24} suggested that ED-3, one of the substructures identified by \citet{dodd:23}, may be associated with Antaeus. We compared the distribution of ED-3 stars listed in \citet{dodd:23} with those of Antaeus and Cluster D in kinematic space, particularly in the $\jphi$--$\jz$ space. Although a few high-$\jz$ stars in Cluster D show dynamical properties broadly similar to those of ED-3 stars ($\langle(\jphi, \jz)\rangle_{\mathrm{ED3}} = (2656.8,\,\,320.3)\kpckms$, $\sigma_{\langle(\jphi, \jz)\rangle_{\mathrm{ED3}}} = (93.1,\,\,46.7)\kpckms$), the median action values of Cluster D and ED-3 differ at the >$3\sigma$ level, indicating that the two substructures are dynamically distinct. Overall, the distributions of Cluster D and Antaeus in dynamical space do not appear to align with that of ED-3, and thus, we tentatively discard the idea that ED-3 is part of Antaeus.

The low mean $\jz$ of Antaeus stars is what makes it stand out from Sequoia. \citet{oria:22} argued that Antaeus shares many similarities in $\etot$, $\jphi$, metallicity distribution, and isochrone age with those for Sequoia, except for its low vertical action. They suggested that Antaeus might lose $\jz$ due to dynamical friction when its progenitor experienced a nearly radial merger \citep[see also][]{amarente:22}, which may also suggest that Antaeus might be a debris core of Sequoia's progenitor. To understand Antaeus's true nature, including the link between Sequoia, Antaeus (and ED-3), a detailed elemental abundance analysis of Antaeus stars is necessary.

\subsubsection{Cluster E: ED-2}\label{sec:ed2}
Cluster E has been found close to Clusters C and D in both IoM and velocity spaces. While it consists of $44$ stars from the initial selections, in the following analysis, we discard $15$ stars, showing a rather larger spread in IoM space (see the bottom-left panel in Figure~\ref{fig:fig4}), by doing $3\sigma$ rejection with the covariance matrix. This process leaves us with $29$ core member stars in Cluster E. Its median metallicity is $-2.19$ dex, making it the most metal-poor structure identified in this study.

The known substructure that is closely associated with Cluster E would be {\it ED-2}, which is a compact dynamical group that has been discovered by \citet{dodd:23}. They used a single linkage-based clustering algorithm that was developed by \citet{lovdal:22} and discovered several new small dynamical groups, including ED-2, that are closely located to Sequoia in IoM space. The structure has been described as a tightly clumped group in velocity space, including three metal-poor stars ($\feh < -1.88$). \citet{balbinot:23} investigated this structure further by using APOGEE DR17 and LAMOST DR8 \citep[including recalibrated values from][]{li:18}. ED-2 is a dynamically compact, cold stellar stream that is close to its pericentre, and it appears to be metal-poor \citep[$\langle\feh\rangle \sim -2.6$;][]{balbinot:23}, of which median metallicity is lower than our measurement (see details below). Due to its low metallicity but cold and compact dynamic nature, they argued that it seems to be one of the progenitor-less substructures, like C-19 \citep{martin:22a, martin:22b, yuan:22} and the Phoenix \citep{balbinot:16, wan:20}. \citet{balbinot:24} updated their analysis by conducting high-resolution spectroscopic analysis of nine stars with the Very Large Telescope (VLT) combined with European Southern Observatory (ESO) archival data for an additional member star and confirmed its globular cluster origin.

The fifth panels of Figure~\ref{fig:fig9} show the distribution of Cluster E (red-filled triangles) and ED-2 (blue-filled circles) in both $\etot - \lz$ (top) and $\vr - \vphi$ (middle) spaces\footnote{Cluster E's member stars found in $95$ per cent confidence ellipses and ED-2 share three stars in common, and their {\it Gaia} DR3 {\tt source\_id}s are the following: 1577319631286280960, 4008274384997266560, and 1578782394069199488.}. In both kinematic spaces, Cluster E seems to share very similar kinematics. Both groups are shown as small overdensities in kinematic spaces, despite a small number of stars (four objects) in Cluster E having positive $\vr$ values. Four stars with negative velocities could imply that Cluster E might experience at least one stream wrap. Given that these stars are only selected from IoM space, however, they could be contaminants sharing similar kinematic properties in IoM space by chance.

In the bottom-right panels of Figure~\ref{fig:fig9}, we compared the metallicity distributions of $17$ ED-2 stars \citep[black histogram;][]{balbinot:23, balbinot:24, ceccarelli:24} and $29$ Cluster E member stars (red histogram). When adopting metallicities from previous studies, we prioritised stars originally identified in \citet{dodd:23} with high-resolution spectroscopic analysis in \citet{balbinot:24}. We also included two additional stars -- {\it Gaia} DR3 1298276602498467072, and 4532592619428218624 -- from \citep{ceccarelli:24} that are not listed in \citet{balbinot:24}. For stars without estimates from high-resolution spectroscopic analysis, we used LAMOST DR8 metallicities of stars originally listed in \citet{dodd:23} as reported in \citet{balbinot:23}. We adopted the original LAMOST DR8 low-resolution metallicities for five stars, except for one star with LAMOST DR8 Mid-resolution results ({\it Gaia} DR3 779616592350301952). 

The metallicity distribution of ED-2 seems to be very metal-poor, which makes it very different from that of Cluster E ($p = 0.010$). The median metallicities of both groups differ, but they still agree within a $1.5\sigma$ level. The difference in metallicity distribution seems to originate either from a systematic offset in the ED-2 member stars from the high-resolution spectroscopic analyses \citep{balbinot:24, ceccarelli:24}, or, less likely, from the possibility that both groups are different. If we consider only six stars from \citet{balbinot:23} with LAMOST DR8 metallicities (mean metallicity of $-2.11^{+0.27}_{-0.10}$) and compare them with Cluster E, they show good agreement within $1\sigma$ level. We tentatively associate Cluster E with ED-2 based on their resemblance shown in kinematic space.

\begin{figure}
    \centering
    \includegraphics{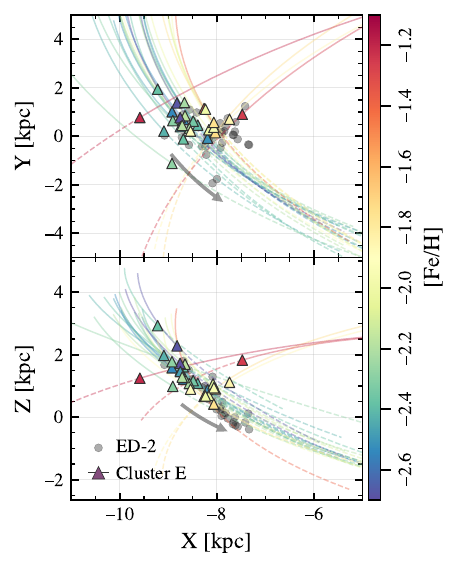}
    \caption{Phase-space distribution of $29$ stars in Cluster E that are found in $95$ per cent confidence ellipse, shown as filled triangles colour-coded by metallicity, in Galactocentric Cartesian coordinates. The $32$ ED-2 member stars from \citet{dodd:23} are shown as grey circles. Solid and dashed lines represent the backwards and forward integrated orbits of Cluster E stars, respectively, and each line is colour-coded by the star's metallicity. Orbits are integrated with a time step of $0.02$ Myr over a total integration time of 40 Myr ($20$ Myr backwards and $20$ Myr forwards). \label{fig:fig11}}
\end{figure}

ED-2 is characterised as a pancake-like structure due to its tight distribution in $X - Z$ plane \citep{balbinot:23}. The distribution of Cluster E's member stars (filled triangles) in Figure~\ref{fig:fig11}, plotted over the original ED-2 member stars (black-filled circles) from \citet{dodd:23}, supports this idea as they show the tight sequence in $X - Z$ plane (bottom panel). Member stars in Cluster E are colour-coded by their metallicities, and most stars close to the ED-2's tight sequence show low metallicities. Coloured lines represent the integrated orbits of our member stars over a total integration time of $40$ million years. Overall, stars show stream-like motion, i.e., showing a narrow spatial distribution in the $X - Z$ plane (bottom panel) with very small velocity dispersion of $V_{X} = 13.5 \pm 3.5 \kms$.

Four stars with positive $V_R$ and higher metallicities ($\feh > -1.6$) show distinct motions compared to the other, more metal-poor member stars of Cluster E. We examined the orbital continuity of these stars with respect to the orbits of the other members with negative $V_R$ by increasing the total integration time to $300$ Myr. The orbital motions (position and velocities) of these four stars are reasonably well-matched to those of other member stars that are forward integrated for $300$ Myr. 

This experiment could be consistent with the idea that Cluster E has experienced a stream wrap, despite the relatively high metallicities of these stars. However, such an interpretation is rather unlikely because the more metal-rich stars are found in the leading component. Given the typical metallicity gradients observed in most galaxies, we would expect more metal-poor stars to be stripped earlier than metal-rich ones. In this case, however, the opposite trend is seen, which makes the stream wrap scenario less likely.

An alternative scenario is that four metal-rich stars originated from a different source rather than from the same progenitor of the metal-poor Cluster E stars. \citet{balbinot:24} suggested that the progenitor of ED-2 may have been an old globular cluster. Given this, these four metal-rich stars might be remnants of the surrounding host dwarf galaxy of ED-2, or, less likely, old accreted stars that were heavily phase-mixed long ago but now happen to share similar dynamic properties. More detailed chemical abundance analyses of these stars will be crucial to deepen our understanding of their true origins.

From this analysis, we conclude that Cluster E is indeed associated with ED-2, despite four stars showing different metallicities. Given the possibility that these stars do not belong to ED-2, we decided to remove them from further analysis to confirm the potential progenitor of each substructure and mark these stars in the full table (see Table~\ref{tab:tab5} and \ref{tab:tab6}) as the potential contaminants of Cluster E.

\subsection{Potential progenitors}\label{sec:progenitor}
The Milky Way halo was formed by multiple merger events and stellar substructures that are considered to be debris of those systems, such as globular clusters or dwarf galaxies, disrupted by the tidal force of the Milky Way \citep[][and references therein]{helmi:20}. With chemical abundance information, it is possible to associate stellar substructures with their potential progenitors. One example is to use the spread in metallicities ($\sigma_{\feh}$). Dwarf galaxies tend to have large metallicity spread ($0.4 - 0.6$) due to their extended star formation history \citep{willman:12, leaman:12}, whereas globular clusters show relatively small intrinsic dispersion in metallicity, implying a short star formation history (with a few exceptions, e.g., $\omega$-Cen, M54, M2, Terzan 5, Liller 1).

To associate substructures identified in this study with potential progenitor systems, we first need to assess DESI's performance in measuring intrinsic metallicity dispersions of dwarf galaxies and globular clusters. Once we have understood this, it will become possible to use the information to trace back the progenitor's object types (i.e., a globular cluster or a dwarf galaxy) of our substructures. Here, we emphasise that our aim is to tentatively link our substructures to the object types of their potential progenitors. Without detailed chemical analyses, connecting overdensities in the IoM space and actual progenitors remains challenging due to the complexity arising in the IoM space \citep[][]{amarente:22, khoperskov:23a, mori:24}. A more detailed investigation on this matter would require a larger sample and detailed elemental abundance information, which will become available in future DESI data releases.

DESI MWS has dedicated observing programmes for dwarf galaxies and globular clusters \citep{cooper:23}, and multiple systems have been observed during the first year of observation. We first obtained the lists of member stars in each globular cluster from \citet{vasiliev:21} and dwarf galaxy from \citet{pace:22a, pace:22b}. The list of globular clusters contains distance \citep{baumgardt:21}, the line-of-sight velocity of each globular cluster from \citet{baumgardt:19}, and proper motions and membership probability computed by \citet{vasiliev:21}. In the case of dwarf galaxies, the list contains the membership probability, distance, proper motions, and the line-of-sight velocity of each system from \citet{pace:22a}. To obtain member stars observed by DESI, we crossmatched those lists to the Y1 catalogue. We included the Sagittarius dwarf galaxy (Sgr) and found \num{13054} member stars in {\it Gaia} EDR3 \citep{ramos:22} that were observed by DESI. To avoid contamination by field stars, we only selected stars with high membership probability (i.e., $P > 0.9$) from \citet{vasiliev:21} for globular clusters, \citet{pace:22a} for dwarf galaxies, and \citet{ramos:22} for Sgr. From this selection, the list consists of $15$ globular clusters and eight dwarf galaxies: this includes stars with an absolute velocity difference, $|V_{\mathrm{rad, star}} - V_{\mathrm{system}}|$, less than $3\sigma$, where $\sigma$ is a quadrature sum of the measurement error in the star's radial velocity ($\sigma_{V_{\mathrm{rad, star}}}$) and the system's velocity dispersion ($\sigma_{V_{\mathrm{system}}}$). 

Among these systems, we only included eight globular clusters (M53: 53 stars, NGC 5053: 25 stars, M3: 39 stars, M5: 290 stars, M13: 153 stars, M12: 15 stars, M92: 60 stars, M2: 26 stars) and three dwarf galaxies (Draco: 97 stars, Sextans: 50 stars, and Sgr: 550 stars), where DESI observed more than $15$ member stars with relatively good signal-to-noise in the $b$ band ($\mathrm{S/N}_{b}> 10$). For globular clusters, we added an additional criterion: \(\feh < 0.0\). The signal-to-noise cut is introduced in order to select stars with better observational quality without too many stars in the substructures being rejected from this cut, i.e., less than $\sim 10$ per cent of stars in each substructure are removed from the newly introduced cut. 

The Metallicity dispersion of each system is a combination of the intrinsic dispersion of the system and DESI's systematic uncertainty in measuring metallicities. We adopted the upper limit of the systematic uncertainty of DESI as approximately $0.1$ dex (Koposov et al., {\it in preparation}) and estimated intrinsic metallicity dispersions ($\sigma_{\feh}$) of eight globular clusters and three dwarf galaxies via maximum log-likelihood analysis. For this purpose, we assumed that the metallicity distribution of each system follows a Gaussian distribution.

\begin{figure}
    \centering
    \includegraphics{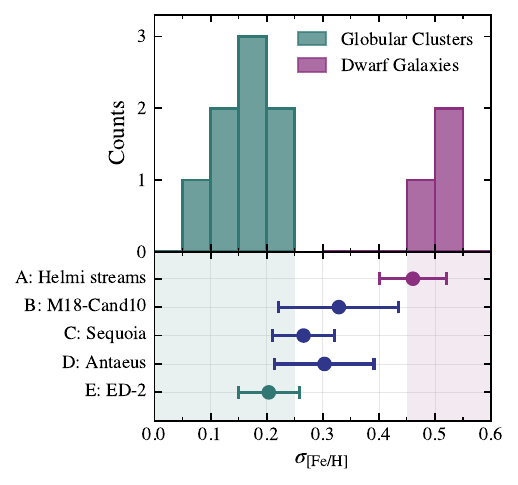}
    \caption{The top and bottom panels show the intrinsic metallicity dispersion of eight globular clusters (green histogram) and three dwarf galaxies (purple histogram) found in the DESI Y1 catalogue and the observed metallicity dispersion of the DESI dynamic groups, respectively. Error bars shown in the bottom panel represent the uncertainty of the metallicity dispersion, which is computed by using simple random sampling with replacements. We associate substructures in the shaded regions with globular clusters (green, $\sigma_{\feh} < 0.25$) or dwarf galaxy origin (purple, $\sigma_{\feh} > 0.45$), while the blue points between the shaded regions are ambiguous and less likely to be associated with their progenitors. \label{fig:fig12}}
\end{figure}

The top panel of Figure~\ref{fig:fig12} shows the metallicity dispersion of eight globular clusters (green histogram) and three dwarf galaxies (purple histogram) that were observed by DESI. The smallest intrinsic metallicity dispersion of dwarf galaxies is $0.49$ dex (Sgr), whereas globular clusters have their intrinsic metallicity spreads ranging from $0.08$ dex (M5) to $0.25$ dex (NGC 5024). From these, we set our threshold of progenitor systems as follows: if the substructure's metallicity dispersion observed by DESI is smaller than $0.25$ dex\footnote{Some globular clusters in DESI Y1 data release still show large intrinsic metallicity dispersions. In fact, Globular clusters have low metallicity dispersion, which may not be detectable by DESI \citep[$\sigma_{\feh} < 0.10$;][and references therein]{bailin:19}.}, the substructure is highly likely to originate from a globular cluster. On the contrary, if the substructure shows a spread in its metallicity distribution larger than $0.45$ dex, then it could be a dwarf galaxy debris.

All substructures detected in this study show metallicity spreads, as shown in the bottom panels in Figures~\ref{fig:fig8} and \ref{fig:fig9}. We obtain the observed metallicity dispersion of each substructure and its uncertainty and show them in the bottom panel in Figure~\ref{fig:fig12}. The metallicity dispersion of each substructure (filled circle) was calculated from the 16th -- 84th percentile dispersion in metallicities of member stars in $95$ per cent confidence ellipse, but also satisfying the signal-to-noise condition of $\mathrm{S/N}_{b}> 10$. We note that the observed metallicity dispersion for each substructure still contains the systematic uncertainty from DESI. Applying this additional condition, the numbers of stars that we obtained for each cluster are $95$ (A: Helmi stream), $22$ (B: M18-Cand10), $27$ (C: Sequoia), $38$ (D: Antaeus), and $23$ (E: ED-2) stars. The uncertainty of the metallicity dispersion (error bar) was computed by using simple random sampling with replacements (i.e., bootstrapping), and we repeated the sampling method \num{10000} times.

Clusters A and E are the ones we can associate with their progenitors at a higher than $2\sigma$ level: Cluster A has a metallicity dispersion of $0.46$ dex with an uncertainty of $0.060$ dex, which makes it most likely that it originated from a dwarf galaxy at a $ 2.5\sigma$ confidence level. Cluster E ($\langle\feh\rangle = 0.20 \pm 0.055$ dex) appears to be a disrupted globular cluster, given that its metallicity dispersion is smaller than $0.25$ dex. Since Cluster A has a substantially larger metallicity dispersion uncertainty, further elemental abundance analyses would be highly beneficial to determine their true origins. The potential progenitors of the other three substructures (Clusters B, C, and D) remain uncertain. Their metallicity dispersion falls between the thresholds of globular clusters or dwarf galaxies. 

Although we ruled out the association of Cluster B with Jhelum, understanding the origin of Jhelum may provide us with a clue about Cluster B. Given their similarities in IoM space and chemistry, they might originate from the same progenitor (see \S~\ref{sec:cluster_b}). While the progenitor type of Jhelum is still debatable, several analyses, including high-resolution spectroscopic analysis by \citet{ji:20}, suggest its dwarf galaxy origin \citep{bonaca:21, li:22}. Cluster B might be debris of the same dwarf galaxy that formed the Jhelum stream, given its relatively large metallicity dispersion ($\sigma_{\feh_{\mathrm{B}}} = 0.33 \pm 0.11$). Cluster B could be the outcome of a close encounter between the progenitor dwarf galaxy of Jhelum and another massive galaxy more recently accreted to the Milky Way. This can explain the similarity of Cluster B with Jhelum in chemistry and most of its dynamic properties. This conclusion is in agreement with the recent result from \citet{woudenberg:23}, reporting the possible interaction between Jhelum and the Sagittarius dwarf galaxy.

Although the metallicity dispersion of Cluster C ($\langle\feh\rangle = 0.27 \pm 0.055$ dex) is greater than the threshold typically associated with globular clusters, the difference remains within $1\sigma$. Meanwhile, its dispersion is also lower than that of typical dwarf galaxies with a deviation from the dwarf galaxy threshold greater than $3\sigma$. This is particularly interesting because Sequoia has been considered a remnant of a massive progenitor \citep{koppelman:19b} due to its widespread presence in $\etot - \lz$ space. One possible explanation for this apparent contradiction could be attributed to our use of HDBSCAN*, which may favour stars that are more tightly clumped in clustering space. We observe that stars with similar kinematics to Cluster C show a similar metallicity distribution but with a wider spread ($\langle \feh \rangle = -1.70 \pm 0.44$; Figure~\ref{fig:fig8}). This suggests that our selection may have only chosen a subset of stars that are relatively more concentrated in dynamical space within a larger structure. This interpretation aligns with the possibility that Cluster C is a fragment of a larger accreted system -- such as a massive dwarf galaxy -- whose stars were stripped over time. In such a scenario, multiple distinct clumps can form in dynamical space as the system is disrupted, and Cluster C may represent one of these substructures. The greater metallicity dispersion of stars sharing similar kinematic properties and the mean metallicity to those of Sequoia suggests that the true nature of Sequoia is a debris of the massive dwarf galaxy.

The other likely explanation is that the stellar population selected from the high energy, retrograde region in $\etot - \lz$ space) consists of member stars from different substructures. This hypothesis was first suggested by \citet{ruiz-lara:22a}, who discovered that the extent of Sequoia is occupied by three different clusters (\#62, \#63, \#64) with low average metallicities. Our clustering results are fairly similar to their conclusion; Given the distribution in $\etot - \lz$ space, Clusters C (Sequoia) and D (Antaeus) appear to share similar kinematic properties with \#62 and \#64. In addition, the average metallicity and its dispersion of Clusters C and D are consistent with \#62 and \#64 at the $1\sigma$ level, even though the mean metallicity of Cluster C is $\sim 0.2$ dex lower than that of \#62. Given the different $\jz$ and low metallicities, Clusters C and D seem to be independent substructures, which opens the possibility that the progenitor of Cluster C might be a system that is less massive than we previously thought. This is also supported by \citet{chen:24}, who argued that major disruptive merger events (e.g., GSE) would significantly impact the smaller progenitors accreted around (or before) the time of the GSE. This would substantially change the orbital actions of these smaller progenitors and make them less recognisable in kinematic spaces. Similarly, substructures that originated from more recently accreted progenitors would be comparatively more easily discovered, as suggested by \citet{wu:22}.

With a $3\sigma$-confidence level, the metallicity dispersion of Cluster E ($0.20 \pm 0.055$ dex) implies that it is a tidally disrupted globular cluster, which is comparable with the latest development in the discussion about the progenitor of ED-2 suggested by \citet{balbinot:24}. Tracing the origin of ED-2 was ambiguous \citep{balbinot:23, ceccarelli:24}, until \citet{balbinot:24} published a high-resolution spectroscopic analysis of $10$ ED-2 member stars with the VLT. They suggested that ED-2 is a low-mass disrupted star cluster with the upper mass limit of $4.2\times10^{4} M_{\odot}$, given its distribution in IoM space and the intrinsic metallicity dispersion, which is close to zero. They found that the age of the progenitor is comparable with that of M92 based on good agreement with the distribution of stars in both ED-2 and M92 in the {\it Gaia} colour-magnitude diagram. 

Our analysis can be significantly improved by constraining better metallicity dispersion of DESI data. With the future DESI MWS data releases, we will also be able to access additional elemental abundance information on C, Mg, and Ca, which will allow us to have a better understanding of their potential progenitors and characteristics. In addition, high-resolution follow-up studies of member stars would complement our analysis, providing a wider range of elemental abundances compared to DESI. Chemical tagging analysis of these stars would be beneficial for gaining valuable insights into their origins and the assembly history of the inner stellar halo. 

\subsection{Limitations}\label{sec:limitations}
Although our approach to identifying substructures has been carefully planned and laid out, a few questions have arisen, mainly due to missing nearby substructures reported from other studies. Here, we briefly address four main limitations in this study. Using more accurate distances would be a solution for most of these questions. 

\subsubsection{HDBSCAN* as a tool for substructure search}
Our findings are mostly phase-mixed but still clustering in kinematic spaces. The optimisation process for the input hyperparameters of HDBSCAN* relies on the distribution of mock datasets generated using a simple assumption about the stellar populations in the Galactic halo (see also, \S~\ref{sec:mockdata}). Our detection depends on optimised hyperparameters, which were tuned to prioritise the search for structures tightly grouped in kinematic spaces (including velocity space). Consequently, using HDBSCAN* exclusively in velocity space requires caution, as the clustering results may only provide highly concentrated substructures with cold progenitors (i.e., a bottom-right panel in Figure~\ref{fig:fig6}) rather than substructures with massive progenitors that would be more spatially coherent in velocity space.

\subsubsection{No detection of GSE or thick disc populations}
Fundamentally, HDBSCAN* can detect an obvious overdensity within a sample distribution even if the hyperparameters — particularly {\tt min\_sample\_size} —  are set smaller than the cluster’s original size. Given this, it might be somewhat surprising to readers that we were not able to detect the most prominent structure in the Galactic halo, the GSE.
    
This outcome is likely the result of our hyperparameter tuning process, which includes the experimental designs for creating the mock dataset. As briefly mentioned in \S~\ref{sec:mockdata}, the GSE-like structure was introduced as part of a global distribution of local halo stars, alongside thick disc and isotropic halo populations. In these mock datasets, the distribution of GSE-like stars is shown as a large-scale structure in dynamic space, but those stars do not resemble a compact, dense cluster with a significantly higher number density than the surrounding area. Unlike stars in $15$ mock clusters, we did not assign labels to mock stars belonging to the thick disc, GSE-like, or isotropic halo structures. These stars, part of large-scale structures, were not identified as separate clusters during the tuning process, and consequently, it is expected that HDBSCAN* under our hyperparameter settings does not identify GSE as a distinct structure in the observational dataset.

Additionally, with DESI data, the distribution of stars in action space (e.g., Figure~\ref{fig:fig6}) may affect the performance of HDBSCAN*. While the GSE may show some level of concentration at lower energy in the $\etot - \lz$ space, it appears as a relatively homogeneous, broadly distributed structure in the logarithmic action space. Its distribution in dynamical space could help explain why HDBSCAN* struggled to detect the GSE as a distinct structure. The algorithm is more likely to identify clusters with higher local density, but the GSE's distribution in these spaces may have led to its classification as part of the general halo population instead of a separate structure.

In fact, when we conducted the clustering analysis on the DESI halo subset, HDBSCAN* identified a single large-scale structure that includes both GSE and thick disc populations. This outcome aligns with the HDBSCAN*'s behaviour, which detects any overdensity larger than the minimum cluster size threshold, {\tt min\_cluster\_size}. However, since this single large-scale structure that we identified in DESI data shows a continuous distribution of the GSE and thick disc population, we intentionally classified those stars as local halo stars.

\subsubsection{Lack of globular clusters}
As we mentioned above, DESI has observed more than $15$ globular clusters during the first year of observation. A few of those clusters (i.e., M5, M12, M13, and M92) are found within the $10$ kpc distance range in the vicinity of the Sun. As such, it raises the question of why HDBSCAN* was not able to detect globular clusters in this study. Although $1,094$ stars in globular clusters are observed by DESI during the first year of the observing period, only nine stars remain in the Y1 halo subset. This enormous difference in the number of globular cluster member stars in the Y1 catalogue and the halo subset appears to be caused by distance estimates and their distance error uncertainties adopted from \citetalias{bailer-jones:21}.
    
More than $95$ per cent of stars in globular clusters provided by the DESI Y1 catalogue have overestimated (or underestimated) distances from \citetalias{bailer-jones:21}. The halo subset in this study contains only nine stars in globular clusters (six stars in M5, two in M12, and one in M92). While the number of stars is not large enough to be detected by HDBSCAN* due to our hyperparameter choice, it is worth noting that \citetalias{bailer-jones:21} distances to stars in globular clusters show huge differences ($>3\sigma_{D_{\mathrm{sys}}}$) from the systems' distances from \citet{baumgardt:21}. \citetalias{bailer-jones:21} acknowledged this issue (i.e., limitation of estimating the distances to the stars in a cluster) as their approach is intended to provide distances for each source independently. As such, the estimated distance to the cluster by averaging \citetalias{bailer-jones:21} distances to the member stars would be a result of a combination of priors that could dominate the distance estimate, especially when the individual member stars have large fractional parallax uncertainties ($\sigma_{\pi}/\pi > 1$) or negative parallaxes. 

Distance error is a major factor in causing dispersion in IoM space; about $12$ - $20$ per cent of distance errors introduce $15$ per cent of dispersion in $\etot,\,\,\lz,\,\,\jr$, and $\jz$. These large distance differences introduce larger scatters of globular cluster stars in IoM space, which makes it more difficult to detect member stars using a density-based clustering algorithm. This issue may become less significant with more precise distances to stars with fainter magnitudes found in crowded fields, which will be provided in future surveys, such as {\it Gaia} DR4.
    
\subsubsection{Missing substructures}
Many studies reported nearby substructures in the Solar neighbourhood by solely using kinematics or chemical tagging, or using the combined information of both kinematics and chemistry. Considering the similarity of our approaches, it raises the question of why we were not able to detect nearby structures, like Kraken/Heracles \citep{massari:19, kruijssen:19, kruijssen:20, horta:21} or Thamnos 1 and 2 \citep{koppelman:19b}, or nearby stellar streams from {\tt STREAMFINDER} \citep{malhan:18a, malhan:18b, ibata:19} in the DESI observing fields, such as Slidr \citep[$d_{\odot} = 2.99$ kpc, $\feh = -1.8$;][]{ibata:19, malhan:22}, Sylgr \citep[$d_{\odot} = 3.77$ kpc, $\feh = -2.92$;][]{ibata:19, malhan:22}, and Gaia-1 \citep[$d_{\odot} = 5.57$ kpc, $\feh = -1.36$;][]{malhan:18b, malhan:22}. 

This issue could be due to the incompleteness of the DESI Y1 catalogue, unlike the analyses that entirely relied on {\it Gaia} data (e.g., {\tt STREAMFINDER}, our study used only a fraction of {\it Gaia} stars with DESI spectra.) Given the DESI MWS observing footprint \citep{cooper:23}, our coverage near the Galactic plane and centre is low, which appears to be a reason that we could not detect the Kraken/Heracles structure. Furthermore, this issue can be linked to different distance estimates adopted in our work and other studies. For example, the DESI Y1 catalogue and a catalogue of $5,960$ stream candidate stars from {\tt STREAMFINDER} \citep{ibata:21} have $480$ stars in common, including the streams mentioned above (Slidr, Sylgr, Gaia-1) and other streams, such as Leiptr, GD-1, Orphan, Ylgr, Pal-5, Sv\"ol, Fj\"orm, Gj\"oll, C-1, M92, Phlegethon, and Hr\'id. Among these, only $16$ stars are included in the halo subset, mainly due to the distance error cut that we introduced in \S~\ref{sec:halo_subset}. Among those excluded from the distance error cut, $180$ stream candidate stars have larger fractional parallax uncertainties ($\sigma_{\pi}/\pi > 1$) or negative parallaxes, which could be the reason for the large uncertainty in \citetalias{bailer-jones:21} distances to those stars. Even if stars have good distance estimates, member stars in Sv\"ol would have been removed due to the circularity cuts that we introduced in \S~\ref{sec:halo_subset}. Updated parallax information from the future {\it Gaia} data releases will provide distance information with smaller uncertainties and will yield a better understanding of their nature.

\subsubsection{Associating overdensities with accretion events}
Understanding the origins of substructures in the Solar neighbourhood is one of the key science topics in Galactic archaeology. These overdensities are often associated with independent accretion events based on their chemodynamic properties (i.e., dynamic characteristics and metallicity distribution function). However, recent studies with Milky Way-like galaxy simulations warn that a cautious approach is required before concluding that all observed overdensities in chemodynamic space correspond to distinct accretion events \citep{jean-baptiste:17, amarente:22, khoperskov:23a, mori:24}. 
    
These simulation results suggest that a single, massive accretion event can create multiple overdensities in kinematic spaces \citet{amarente:22, mori:24} and that stars with different metallicities originating from the same accretion event could be found in different regions of IoM space \citet{mori:24}. \citet{khoperskov:23a} also suggested that major merger events could introduce heating and redistribute the pre-existing stellar populations. Furthermore, different accretion events can form overdensities that substantially overlap in kinematic spaces \citep{simpson:19, kizhuprakkat:24}, which makes it even more complicated to associate overdensities in chemodynamic space with accretion events.

Our analysis in \S~\ref{sec:progenitor} has attempted to trace substructures to the object types of their origins, but we remain careful in our interpretations, considering the findings of other studies. A more detailed analysis, incorporating additional information such as elemental abundances, is necessary to fully understand the origins of these substructures.\\

\section{Summary and Conclusion}\label{sec:summary}
We present five stellar substructures in the local Galactic halo identified in a subset of $138,861$ nearby ($d < 5$ kpc) stars in the Galactic thick disc and halo in the DESI MWS Y1 catalogue (see \S~\ref{sec:data}). Five dynamic groups (Clusters A - E) were identified in IoM and velocity spaces by using one of the density-based unsupervised clustering algorithms, HDBSCAN* (see \S~\ref{sec:results}). Given the metallicity distribution of each substructure, we found that Clusters A, B, and E are chemically distinct from groups of stars sharing similar kinematics with them, whereas Clusters C and D do not show clear differences in metal abundances with respect to background stars with similar kinematics (see \S~\ref{sec:validation}). 

All five clusters appear to associate with the ones previously reported in various studies, including the Helmi streams (Cluster A), M18-Cand10/MMH-1 (Cluster B), Sequoia (Cluster C), Antaeus (Cluster D), and ED-2 (Cluster E). The DESI metallicity distribution of each kinematic group is in good agreement with that from various external spectroscopic surveys (see \S~\ref{sec:association}). Computing metallicity dispersion and its uncertainty, we tentatively associated Cluster A with a dwarf galaxy origin and Cluster E with a globular cluster origin, whereas the origins of the other three dynamic groups (Clusters B, C, and D) remain unclear (see \S~\ref{sec:progenitor}). A more detailed spectroscopic analysis with various elemental abundances would be required to trace them back to their progenitors.

Although our use of HDBSCAN* to identify substructures in the local Galactic halo did not lead to the discovery of new substructures, our search results confirm the presence of known substructures in the DESI Y1 data, validating the reliability of using HDBSCAN* for such searches. The application of unsupervised clustering algorithms enables substructure searches without prior knowledge, minimising selection bias. Using this method, we may recover additional known substructures in the upcoming DESI MWS catalogues, and any newly discovered substructures are likely to be genuine detections.

Of course, our approach to searching for substructures has some caveats and misses many known stellar objects and substructures, including globular clusters and streams (see \S~\ref{sec:limitations}). While these issues could have been mitigated by relaxing clustering conditions (i.e., using smaller {\tt min\_cluster\_size} and/or {\tt min\_samples}, as well as {\tt leaf} clustering), future data releases from DESI will resolve some of the issues raised in this work. Future datasets will include radial velocities and metallicities of newly observed objects, including more member stars in known globular clusters and dwarf galaxies, and stars with better signal-to-noise due to several revisits\footnote{Although DESI will visit {\tt MAIN} targets only once, some MWS targets may be observed more than once if no objects, including MWS targets that have been observed or objects from other programmes, are available to a fibre in a tile \citep{cooper:23}.}, allowing better constraints on stellar parameters. 

The ongoing development of the SP pipeline will provide more precise estimates of elemental abundances of C, Mg, and Ca. This will offer opportunities to conduct clustering in elemental abundance spaces and to have a better understanding of the chemical enrichment history of the Milky Way stellar halo. The synergy between future DESI data releases and more precisely constrained astrometric parameters (distance and proper motion) provided by the upcoming 4th {\it Gaia} data release will boost our exploration of hidden substructures in the Galactic halo.

\section*{Acknowledgements}
The authors are grateful to our referee, Dr Davide Massari, for the constructive comments, which helped greatly improve the content of this paper. B.K. thanks Prof Ting S. Li, Dr GyuChul Myeong, and Prof Paul J. McMillan for sharing their datasets. S.E.K. acknowledges support from the Science \& Technology Facilities Council (STFC) grant ST/Y001001/1. T.S.L. acknowledges financial support from Natural Sciences and Engineering Research Council of Canada (NSERC) through grant RGPIN-2022-04794. A.P.C.\ acknowledges support from the Taiwanese Ministry of Education Yushan Fellowship (MOE-113-YSFMS-0002-001-P2) and the Taiwanese National Science and Technology Council (113-2112-M-007-009). The authors thank the two DESI internal reviewers (Dr Alexander H. Riley and Jiwon Jesse Han) and the DESI Publication Board handler (Dr Siyi Xu).

This material is based upon work supported by the U.S. Department of Energy (DOE), Office of Science, Office of High-Energy Physics, under Contract No. DE–AC02–05CH11231, and by the National Energy Research Scientific Computing Center, a DOE Office of Science User Facility under the same contract. Additional support for DESI was provided by the U.S. National Science Foundation (NSF), Division of Astronomical Sciences under Contract No. AST-0950945 to the NSF’s National Optical-Infrared Astronomy Research Laboratory; the Science and Technology Facilities Council of the United Kingdom; the Gordon and Betty Moore Foundation; the Heising-Simons Foundation; the French Alternative Energies and Atomic Energy Commission (CEA); the National Council of Humanities, Science and Technology of Mexico (CONAHCYT); the Ministry of Science, Innovation and Universities of Spain (MICIU/AEI/10.13039/501100011033), and by the DESI Member Institutions: \url{https://www.desi.lbl.gov/collaborating-institutions}. Any opinions, findings, and conclusions or recommendations expressed in this material are those of the author(s) and do not necessarily reflect the views of the U. S. National Science Foundation, the U. S. Department of Energy, or any of the listed funding agencies.

The authors are honored to be permitted to conduct scientific research on Iolkam Du’ag (Kitt Peak), a mountain with particular significance to the Tohono O’odham Nation.

This work has made use of data from the European Space Agency (ESA) mission {\it Gaia} (\url{https://www.cosmos.esa.int/gaia}), processed by the {\it Gaia} Data Processing and Analysis Consortium (DPAC,
\url{https://www.cosmos.esa.int/web/gaia/dpac/consortium}). Funding for the DPAC has been provided by national institutions, in particular the institutions participating in the {\it Gaia} Multilateral Agreement.

This paper made use of the Whole Sky Database (wsdb) created and maintained by Sergey Koposov at the Institute of Astronomy, Cambridge with financial support from the Science \& Technology Facilities Council (STFC) and the European Research Council (ERC). This publication made use of sqlutilpy v.0.20.0\footnote{\url{https://zenodo.org/doi/10.5281/zenodo.5160118}} \citep{sqlutilpy}.

Funding for the Sloan Digital Sky Survey IV has been provided by the Alfred P. Sloan Foundation, the U.S. Department of Energy Office of Science, and the Participating Institutions. 

SDSS-IV acknowledges support and resources from the Center for High Performance Computing at the University of Utah. The SDSS website is www.sdss4.org.

SDSS-IV is managed by the Astrophysical Research Consortium for the Participating Institutions of the SDSS Collaboration including the Brazilian Participation Group, the Carnegie Institution for Science, Carnegie Mellon University, Center for Astrophysics | Harvard \& Smithsonian, the Chilean Participation Group, the French Participation Group, Instituto de Astrof\'isica de Canarias, The Johns Hopkins University, Kavli Institute for the Physics and Mathematics of the Universe (IPMU) / University of Tokyo, the Korean Participation Group, Lawrence Berkeley National Laboratory, Leibniz Institut f\"ur Astrophysik Potsdam (AIP),  Max-Planck-Institut f\"ur Astronomie (MPIA Heidelberg), Max-Planck-Institut f\"ur Astrophysik (MPA Garching), Max-Planck-Institut f\"ur Extraterrestrische Physik (MPE), National Astronomical Observatories of China, New Mexico State University, New York University, University of Notre Dame, Observat\'ario Nacional / MCTI, The Ohio State University, Pennsylvania State University, Shanghai Astronomical Observatory, United Kingdom Participation Group, Universidad Nacional Aut\'onoma de M\'exico, University of Arizona, University of Colorado Boulder, University of Oxford, University of Portsmouth, University of Utah, University of Virginia, University of Washington, University of Wisconsin, Vanderbilt University, and Yale University.

Guoshoujing Telescope (the Large Sky Area Multi-Object Fiber Spectroscopic Telescope LAMOST) is a National Major Scientific Project built by the Chinese Academy of Sciences. Funding for the project has been provided by the National Development and Reform Commission. LAMOST is operated and managed by the National Astronomical Observatories, Chinese Academy of Sciences.

Based on data acquired at the Anglo-Australian Telescope. We acknowledge the traditional owners of the land on which the AAT stands, the Gamilaraay people, and pay our respects to elders past and present.

This work made use of Astropy:\footnote{\url{http://www.astropy.org}} a community-developed core Python package and an ecosystem of tools and resources for astronomy \citep{astropy:13, astropy:18, astropy:22}.

%%%%%%%%%%%%%%%%%%%%%%%%%%%%%%%%%%%%%%%%%%%%%%%%%%
\section*{Data Availability}
The data used in this analysis is publicly available at \url{https://data.desi.lbl.gov/doc/releases/dr1/vac/mws/}. A full machine-readable table of \num{314} substructure member stars (Table~\ref{tab:tab6}) and data points for reproducing all the figures are available on Zenodo (\url{https://doi.org/10.5281/zenodo.15068958}). 

\section*{Software}
This work made use of the following software dependencies: \texttt{Python v.3.9.13} \citep{python}, \texttt{Jupyter v.3.5.0} \citep{jupyter}, \texttt{Numpy v.1.23.3} \citep{numpy}, \texttt{Scipy v.1.9.3} \citep{scipy}, \texttt{matplotlib v.3.5.2} \citep{matplotlib}, \texttt{scikit-learn v.1.3.0} \citep{scikit-learn}, \texttt{Astropy v.5.1} \citep{astropy:13, astropy:18, astropy:22}, \texttt{Gala v.1.5} \citep{gala, gala_v15}, \texttt{galpy v.1.8.0} \citep{bovy:15}, \texttt{HDBSCAN* v.0.8.33} \citep{hdbscan}, \texttt{sqlutilpy v.0.20.0} \citep{sqlutilpy}.



\bibliographystyle{mnras}
\bibliography{bibilography}

%%%%%%%%%%%%%%%%%%%%%%%%%%%%%%%%%%%%%%%%%%%%%%%%%%

%%%%%%%%%%%%%%%%% APPENDICES %%%%%%%%%%%%%%%%%%%%%

\appendix
\section{Author affiliations}\label{sec:affiliations}
$^{1}$ Institute for Astronomy, University of Edinburgh, Royal Observatory, Blackford Hill, Edinburgh EH9 3HJ, UK\\
$^{2}$ Institute of Astronomy, University of Cambridge, Madingley Road, Cambridge CB3 0HA, UK\\
$^{3}$ Department of Astronomy \& Astrophysics, University of Toronto, Toronto, ON M5S 3H4, Canada\\
$^{4}$ Institute for Computational Cosmology, Department of Physics, Durham University, South Road, Durham DH1 3LE, UK\\
$^{5}$ Institute of Astronomy and Department of Physics, National Tsing Hua University, 101 Kuang-Fu Rd. Sec. 2, Hsinchu 30013, Taiwan\\
$^{6}$ Center for Informatics and Computation in Astronomy, NTHU, 101 Kuang-Fu Rd. Sec. 2, Hsinchu 30013, Taiwan\\
$^{7}$ Department of Physics and Astronomy, Carleton College, One North College Street, Northfield, MN 55057, USA\\
$^{8}$ Department of Astronomy, University of Michigan, Ann Arbor, MI 48109, USA\\
$^{9}$ Center for Astrophysics $|$ Harvard \& Smithsonian, 60 Garden Street, Cambridge, MA 02138, USA \\
$^{10}$ Lawrence Berkeley National Laboratory, 1 Cyclotron Road, Berkeley, CA 94720, USA\\
$^{11}$ Physics Dept., Boston University, 590 Commonwealth Avenue, Boston, MA 02215, USA\\
$^{12}$ Steward Observatory, University of Arizona, 933 N, Cherry Ave, Tucson, AZ 85721, USA\\
$^{13}$ Dipartimento di Fisica ``Aldo Pontremoli'', Universit\`a degli Studi di Milano, Via Celoria 16, I-20133 Milano, Italy \\
$^{14}$ Department of Physics \& Astronomy, University College London, Gower Street, London, WC1E 6BT, UK\\
$^{15}$ Department of Physics and Astronomy, The University of Utah, 115 South 1400 East, Salt Lake City, UT 84112, USA\\
$^{16}$ Instituto de F\'{\i}sica, Universidad Nacional Aut\'{o}noma de M\'{e}xico, Cd. de M\'{e}xico C.P. 04510, M\'{e}xico \\
$^{17}$ Departamento de F\'isica, Universidad de los Andes, Cra. 1 No. 18A-10, Edificio Ip, CP 111711, Bogot\'a, Colombia \\
$^{18}$ Observatorio Astron\'omico, Universidad de los Andes, Cra. 1 No. 18A-10, Edificio H, CP 111711 Bogot\'a, Colombia\\
$^{19}$ University of Michigan, Ann Arbor, MI 48109, USA\\
$^{20}$ Fermi National Accelerator Laboratory, PO Box 500, Batavia, IL 60510, USA\\
$^{21}$ Center for Cosmology and AstroParticle Physics, The Ohio State University, 191 West Woodruff Avenue, Columbus, OH 43210, USA\\
$^{22}$ Department of Physics, The Ohio State University, 191 West Woodruff Avenue, Columbus, OH 43210, USA\\
$^{23}$ The Ohio State University, Columbus, 43210 OH, USA\\
$^{24}$ Department of Physics, Southern Methodist University, 3215 Daniel Avenue, Dallas, TX 75275, USA\\
$^{25}$ Sorbonne Universit\'{e}, CNRS/IN2P3, Laboratoire de Physique Nucl\'{e}aire et de Hautes Energies (LPNHE), FR-75005 Paris, France\\
$^{26}$ NSF NOIRLab, 950 N. Cherry Ave., Tucson, AZ 85719, USA\\
$^{27}$ Instituci\'{o} Catalana de Recerca i Estudis Avan\c{c}ats, Passeig de Llu\'{\i}s Companys, 23, 08010 Barcelona, Spain\\
$^{28}$ Institut de F\'{i}sica d'Altes Energies (IFAE), The Barcelona Institute of Science and Technology, Campus UAB, 08193 Bellaterra Barcelona, Spain\\
$^{29}$ Department of Physics and Astronomy, Siena College, 515 Loudon Road, Loudonville, NY 12211, USA\\
$^{30}$ Space Sciences Laboratory, University of California, Berkeley, 7 Gauss Way, Berkeley, CA 94720, USA\\
$^{31}$ University of California, Berkeley, 110 Sproul Hall \#5800 Berkeley, CA 94720, USA\\
$^{32}$ Instituto de Astrof\'{i}sica de Andaluc\'{i}a (CSIC), Glorieta de la Astronom\'{i}a, s/n, E-18008 Granada, Spain\\
$^{33}$ Department of Physics and Astronomy, Sejong University, Seoul, 143-747, Korea\\
$^{34}$ CIEMAT, Avenida Complutense 40, E-28040 Madrid, Spain\\
$^{35}$ Department of Physics, University of Michigan, Ann Arbor, MI 48109, USA\\
$^{36}$ Centre for Advanced Instrumentation, Department of Physics, Durham University, South Road, Durham DH1 3LE, UK\\
$^{37}$ National Astronomical Observatories, Chinese Academy of Sciences, A20 Datun Rd., Chaoyang District, Beijing, 100012, P.R. China\\
%%%%%%%%%%%%%%%%%%%%%%%%%%%%%%%%%%%%%%%%%%%%%%%%%%

% Don't change these lines
\bsp	% typesetting comment
\label{lastpage}
\end{document}